\documentclass{aa}
\usepackage{graphicx, graphics}
\usepackage{CJK}
\usepackage{bbm}
\usepackage{amsmath, amssymb, bm}
\usepackage[utf8]{inputenc}
\usepackage[normalem]{ulem}
\usepackage[nodayofweek]{datetime}

\input{prepcitationformat.txt}
\usepackage{txfonts}
\usepackage{colortbl}

\usepackage{pdflscape}
\usepackage{rotating}

\usepackage[switch]{lineno}

\usepackage{tikzsymbols}
\usepackage{tikz}    
\newcommand{\ExternalLinkSmall}{\!%
    \tikz[x=0.9ex, y=0.9ex, baseline=-0.9ex, blue]{%
        \begin{scope}[x=0.7ex, y=0.7ex]
            \clip (-0.1,-0.1) 
                --++ (-0, 1.2) 
                --++ (0.6, 0) 
                --++ (0, -0.6) 
                --++ (0.6, 0) 
                --++ (0, -1);
            \path[draw, 
                line width = 1, 
                rounded corners=1] 
                (0,0) rectangle (1,1);
        \end{scope}
        \path[draw, line width = 1] (0.5, 0.5) 
            -- (1, 1);
        \path[draw, line width = 1] (0.6, 1) 
            -- (1, 1) -- (1, 0.6);
        }\!\!\!
    }

\definecolor{lime}{HTML}{A6CE39}
\DeclareRobustCommand{\orcidicon}{%
    \raisebox{-3pt}{\begin{tikzpicture}
    \filldraw [lime, yshift=-2pt] (0, 0) circle [radius=0.16]
    node[white] {\raisebox{1pt}{\hspace{0.5pt}\fontfamily{qag}\selectfont\tiny i\scalebox{0.8}{D}}};
    \end{tikzpicture}}
    \hspace{-2.5mm}
    \vspace{-0.25pt}
}

\newcommand*{\QED}{\hfill\ensuremath{\blacksquare}}%
\newcommand{\Qphi}{$\mathcal{Q}_\phi$}
\newcommand{\Uphi}{$\mathcal{U}_\phi$}
\newcommand{\Itot}{$\mathcal{I}_{\rm tot}$}

\newdateformat{monthyearday}{%
  \THEYEAR\ \monthname[\THEMONTH] \twodigit{\THEDAY}}
\newdateformat{daymonthyear}{%
  \twodigit{\THEDAY}\ \monthname[\THEMONTH] \THEYEAR}
\newcommand\numberthis{\addtocounter{equation}{1}\tag{\theequation}}

\newcommand{\programESO}[1]{#1~\href{http://archive.eso.org/wdb/wdb/eso/sched_rep_arc/query?progid=#1}{\ExternalLinkSmall}}

\newcommand{\orcidauthor}[2]{#2\href{http://orcid.org/#1}{\orcidicon}}

\titlerunning{SPHERE star-hopping Disks}
\authorrunning{Ren et al.}

\begin{document}
\begin{CJK*}{UTF8}{gbsn}
\title{Protoplanetary disks in $K_s$-band total intensity and polarized light\thanks{Based on observations performed with VLT/SPHERE under program ID 0103.C-0470, 105.209E, 105.20HV, 106.21HJ, 108.22EE, 105.20JB.}
}

\author{
\orcidauthor{0000-0003-1698-9696}{Bin B. Ren (任彬)}\thanks{Marie Sk\l odowska-Curie Fellow}\inst{\ref{inst-oca}, \ref{inst-uga}, \ref{inst-cit}} 
\and
\orcidauthor{0000-0002-7695-7605}{Myriam Benisty}\inst{\ref{inst-oca}, \ref{inst-uga}} 
\and
\orcidauthor{0000-0002-4438-1971}{Christian Ginski}\inst{\ref{inst-galway}}
\and
\orcidauthor{0000-0003-1451-6836}{Ryo Tazaki}\inst{\ref{inst-uga}} 
\and
\orcidauthor{0000-0003-0354-0187}{Nicole L. Wallack}\inst{\ref{inst-epl}}
\and
\orcidauthor{0000-0001-9325-2511}{Julien Milli}\inst{\ref{inst-uga}}
\and
\orcidauthor{0000-0002-4266-0643}{Antonio Garufi}\inst{\ref{inst-inaf}}
\and
\orcidauthor{0000-0001-7258-770X}{Jaehan Bae}\inst{\ref{inst-ufl}}
\and
\orcidauthor{0000-0003-4689-2684}{Stefano Facchini}\inst{\ref{inst-milano}}
\and
\orcidauthor{0000-0002-1637-7393}{Fran\c{c}ois M\'enard}\inst{\ref{inst-uga}}
\and
\orcidauthor{0000-0001-8764-1780}{Paola Pinilla}\inst{\ref{inst-ucl}}
\and
\orcidauthor{0000-0003-1371-8890}{C. Swastik}\inst{\ref{inst-iia}} 
\and
\orcidauthor{000-0003-1534-5186}{Richard Teague}\inst{\ref{inst-mit}}
\and
\orcidauthor{0000-0001-8269-324X}{Zahed Wahhaj}\inst{\ref{inst-eso}}
}

\institute{
Universit\'{e} C\^{o}te d'Azur, Observatoire de la C\^{o}te d'Azur, CNRS, Laboratoire Lagrange, Bd de l'Observatoire, CS 34229, 06304 Nice cedex 4, France; 
\url{bin.ren@oca.eu} \label{inst-oca}
\and
Universit\'{e} Grenoble Alpes, CNRS, Institut de Plan\'{e}tologie et d'Astrophysique (IPAG), F-38000 Grenoble, France \label{inst-uga}
\and
Department of Astronomy, California Institute of Technology, MC 249-17, 1200 E California Blvd, Pasadena, CA 91125, USA \label{inst-cit}
\and
School of Natural Sciences, University of Galway, University Road, H91 TK33 Galway, Ireland \label{inst-galway}
 \and
 Earth and Planets Laboratory, Carnegie Institution for Science, Washington, DC 20015, USA \label{inst-epl}
\and
INAF, Osservatorio Astrofisico di Arcetri, Largo Enrico Fermi 5, I-50125 Firenze, Italy \label{inst-inaf}
\and
Department of Astronomy, University of Florida, Gainesville, FL 32611, USA \label{inst-ufl}
\and
Dipartimento di Fisica, Universit\'{a} degli Studi di Milano, via Celoria 16, I-20133 Milano, Italy \label{inst-milano}
\and
Mullard Space Science Laboratory, University College London, Holmbury St Mary, Dorking, Surrey RH5 6NT, UK \label{inst-ucl}
\and
Indian Institute of Astrophysics, Koramangala 2nd Block, Bangalore 560034, India \label{inst-iia}
\and
Department of Earth, Atmospheric, and Planetary Sciences, Massachusetts Institute of Technology, Cambridge, MA 02139, USA \label{inst-mit}
\and
European Southern Observatory, Alonso de C\'ordova 3107, Vitacura Casilla 19001, Santiago, Chile \label{inst-eso}
}

\date{Received 04 July 2023 / Revised 12 October 2023 / Accepted 12 October 2023} 

\abstract
{Diverse morphology in protoplanetary disks can result from planet-disk interaction, suggesting the presence of forming planets. Characterizing disks can inform the formation environments of planets. To date, most imaging campaigns have probed the polarized light from disks, which is only a fraction of the total scattered light and not very sensitive to planetary emission.}
{We aim to observe and characterize protoplanetary disk systems in the near-infrared in both polarized and total intensity light,
to carry out an unprecedented study of the dust scattering properties of disks, as well as of any possible planetary companions.}
{Using the star-hopping mode of the SPHERE instrument at the Very Large Telescope, we observed 29 young stars hosting protoplanetary disks and their reference stars in the $K_s$-band polarized light. We extracted disk signals in total intensity by removing stellar light using the corresponding reference star observations, by adopting the data imputation concept with sequential non-negative matrix factorization (DI-sNMF). For well-recovered disks in both polarized and total intensity light, we parameterized the polarization fraction phase functions using scaled beta distribution. We investigated the empirical DI-sNMF detectability of disks using logistic regression. For systems with SPHERE data in $Y$-/$J$-/$H$-band, we summarized their polarized color at ${\approx}90^\circ$ scattering angle.}
{We obtained high-quality disk images in total intensity for 15 systems and in polarized light for 23 systems. Total intensity detectability of disks primarily depends on host star brightness, which determines adaptive-optics control ring imagery and thus stellar signals capture using DI-sNMF. The peak of polarization fraction tentatively correlates with the peak scattering angle, which could be reproduced using certain composition for compact dust, yet more detailed modeling studies are needed. Most of disks are blue in polarized $J-K_s$ color, and the fact that they are relatively redder as stellar luminosity increases indicates larger scatterers.}
{High-quality disk imagery in both total intensity and polarized light allows for disk characterization in polarization fraction. The combination of them reduces the confusion between disk and planetary signals.}


\maketitle

\section{Introduction}

In the past 10 years, the advent of high angular resolution facilities enabled the detection of numerous disk substructures, such as rings, spirals, dust-depleted cavities, in the near-infrared scattered light \citep[e.g.,][]{benisty15, Benisty2023, wagner18, Shuai2022} and in the (sub-)millimeter/mm regime \citep[e.g.,][]{Francis20, Long2022}, indicating the ubiquity of substructures in large, bright disks \citep{Bae2023}. These substructures can be interpreted as evidence of planet-disk interactions, suggesting the presence of an underlying yet-undetected population of young exoplanets \citep[e.g.,][]{Dong2012}. Additional support for this interpretation recently came from the detection of local velocity deviations in the gaseous outer disk velocity field probed with ALMA \citep[e.g.,][]{Pinte2018, Teague2018, Pinte2020, woelfer23, Stadler2023}. Scattered light surveys also pointed out a large fraction of infrared-faint disks, that appear more compact and featureless in scattered light because of self-shadowing effects \citep[e.g.,][]{Garufi2022}. These disks however often host substructures in the sub-millimeter \citep[e.g.,][]{Long2018} that could be due to planets.

The presence of massive planets inside cavities was also suggested in transition disks (disks with depleted inner cavities; \citealp{Bae2019}) and confirmed in at least one system, PDS~70, with the detection of two protoplanets \citep{Keppler2018, Haffert2019}. The range of plausible mass for the companion(s) in these disks is however quite large, as eccentric stellar companion could be sculpting the cavity \citep[e.g.,][]{Calcino2019} as found in the HD~142527 system \citep{Balmer2022}. In that specific case, the companion is also leading to a misaligned inner disk, which casts a shadow on the outer disk \citep{Price2018}. Such misalignments were found in at least 6 transition disks \citep{bohn22}. Whether these features are of planetary or stellar nature, the search for the perturbers, which are responsible for all the observed disk substructures \citep[e.g.,][]{AsensioTorres2021, Cugno2023}, is of prime importance to understand the formation and evolution of planetary systems. The detection of these perturbers would offer crucial observational evidence to test planet-disk interaction theories \citep[e.g.,][]{Dong2015} and constrain the overall evolution of a planetary system \citep{Bae2019}. However, directly imaging planets embedded in bright and highly structured disks is very challenging with current instruments. Until now, all claims but PDS~70 still require confirmation \citep[e.g.,][]{Kraus2012, Sallum2015, Quanz2015, Reggiani2018, Wagner2019, Boccaletti2020, Uyama2020, Currie22a, Hammond2023, Law2023, Wagner2023}. 

To observe exoplanetary systems with high-contrast imaging, observation strategies including angular differential imaging (ADI; \citealp{Marois2006}, where parallactic angle diversity of observations is used to remove star light) have enabled the detection of prototypical planetary systems (e.g., HR~8799; \citealp{Marois2008}). Nevertheless, ADI detections are still limited by self-subtraction at close-in regions from the stars \citep[e.g.,][]{milli12, wahhaj21}, yet these regions are where giant planets are expected to have the most occurrence \citep[1--10~au; from a combination of radial velocity and high-contrast imaging surveys, e.g.,][]{Nielsen2019, Fulton2021}. To overcome this limitation, on the one hand, better optimized post-processing methods for ADI datasets were developed \citep[e.g.,][]{Pairet2021, Flasseur2021, Juillard2022, Juillard2023}. On the other hand, the diversity in archival observational data can enable the usage of other stars as the templates to remove star light and speckles with the reference differential imaging (RDI) data reduction strategy \citep[e.g.,][]{Ruane2019, xie22}. Moving forward along the direction of RDI, the Spectro-Polarimetic High contrast imager for Exoplanets REsearch \citep[SPHERE;][]{Beuzit19} at the Very Large Telescope (VLT) from European Southern Observatory (ESO) initiated the star-hopping mode \citep{wahhaj21}, which offers quasi-simultaneous observations of a science star and its reference star, unleashing the full potential in exoplanet imaging in close-in regions for SPHERE. 

Determining dust properties is of fundamental importance for the early stage of grain growth and planetesimal formation, as they will determine the efficiency of grain sticking and fragmentation \citep{Birnstiel2012}. In addition to the planet imaging capabilities with SPHERE, the star-hopping mode enables optimized extraction of disks in scattered light in total intensity. This goes beyond the polarimetric surveys that have been routinely carried out in the near-infrared \citep[e.g.,][]{Avenhaus18, Garufi2020, Ginski2020}, and allows us to better study spatial distribution and properties of dust in the disk \citep[e.g.,][]{Olofsson2023}. With the observations taken in dual-polarimetry imaging \citep[DPI:][]{Langlois10} mode, which probes polarized signals in the scattered light, star-hopping can also offer total intensity imaging from RDI. The combination of both can yield an estimate of the polarization fraction, and thus to better constrain dust properties \citep[e.g., shape, composition:][]{Ginski2023, Tazaki23}. 

In this study, we present the first large survey of protoplanetary disks in total intensity from the ground. As many as 29 young stars are surveyed in $K_s$-band  with VLT/SPHERE in the star-hopping mode. Our target sample consists of both transition disk systems to search for protoplanets that can potentially reside in the close-in regions with star-hopping that are otherwise unachievable \citep{wahhaj21}, and non-transition disk sample of faint disks in the infrared to search for planets in their outer disk regions. We also aim to derive the polarization fraction whenever possible. The paper is structured as follows: Sect.~\ref{sec:obs} provides the description of the observations and data reduction procedure, Sect.~\ref{sec:analysis} presents the polarized light and total intensity maps, Sect.~\ref{sec:contrast} shows the detection limits of companions, and in Sect.~\ref{sec:pol-frac} we present the polarization fraction maps. We summarize and conclude the study in Sect.~\ref{sec:summary}. 

\section{Observations and Data Reduction}
\label{sec:obs}
\subsection{Sample of protoplanetary disks}
The sample analyzed in this work includes 29 young stars from both Herbig AeBe and T Tau stars. These are 13 sources from the Taurus star-forming region (CI Tau, CQ Tau, CY Tau, DL Tau, DM Tau, DN Tau, DS Tau, GM Aur, HD 31648, IP Tau, IQ Tau, LkCa 15, MWC 758), 5 from the Scorpius-Centaurus association (HD 100453, HD 100546, HD 143006, HD 169142, SAO~206462), 3 from Chamaeleon (HD 97048, SZ Cha, SY Cha), 3 from Orion (HD 34282, PDS 201, V1247 Ori), and one from each of the following regions: Lupus (V1094 Sco), $\epsilon$ Cha (PDS 66), $\rho$ Ophiuchus (EM* SR 20), and Perseus (LkHa 330). There is also one isolated source (HD 163296). To ensure the identifiability of these targets, who often have different names from various database, we also list the SIMBAD identifiers \citep{simbad} for these targets in Table~\ref{tab:prop}.

To carry out this survey of protoplanetary disks in total intensity, we selected transition disks with known substructures, previously observed in scattered light, as well as non-transition disks in Taurus, with $R$-band magnitude within the SPHERE limits. Using the \textit{Gaia}~DR3 distances \citep{GaiaDR3} and following the approach in \citet{Garufi2018}, we uniformly calculated the stellar properties for the entire sample. In particular, we retrieved the effective temperature $T_{\rm eff}$ from VizieR \citep{ochsenbein00} and the photometry using the VizieR photometry tool.\footnote{\url{http://vizier.cds.unistra.fr/vizier/sed/}} We calculated the stellar luminosity $L_{\rm star}$ through a PHOENIX model \citep{Hauschildt1999} scaled to the de-reddened brightness in $V$-band. From $T_{\rm eff}$ and $L_{\rm star}$, we derived the interval of values for the stellar 
 age identified by different sets of pre-main-sequence tracks \citep{Siess2000, Bressan2012, Baraffe2015, Choi2016}. We summarize all the derived stellar properties\footnote{The derived uncertainties in this paper are $1\sigma$ (frequentist) or $68\%$ credible intervals (Bayesian) unless otherwise specified.} in Table~\ref{tab:prop}. Our sample has a wide, uniform coverage of both $T_{\rm eff}$ and $L_{\rm star}$, and therefore of $M_{\rm star}$ and system age, with stellar mass from 0.5 $\rm M_{\odot}$ (IQ Tau) to 2.7 $\rm M_{\odot}$ (LkHa 330) and age from ${\sim}1$ Myr (a few Taurus sources) to ${\gtrsim}10$ Myr (e.g., MWC 758).  
 
\begin{table*}
\setlength{\tabcolsep}{1.5pt}
\caption{System parameters and star-hopping observation log \label{tab:prop}}
\hspace{-1cm}\begin{tabular}{c r c r r r r c c c c c r r r}    \hline\hline
id                             & Host       & Name            & Sp     & Sp     & Region       & $d$                       & $L_{\rm star}$            & $M_{\rm star}$ & Age        & Date       & DIT  & $t^*_{\rm exp}$ &            Reference Star & $t_{\rm exp}$\\
                               &            & (SIMBAD)        & Type   & Ref    &              & (pc)                      & ($L_\odot$)               & $(M_\odot)$ & (Myr)      & UTC        & (s)  &           (s) &                           &           (s)\\
(1)                            & (2)        & (3)             & (4)    & (5)    & (6)          & (7)                       & (8)                       & (9)        & (10)  & (11)       & (12) &           (13) &                      (14) &            (15)\\ \hline 
\rowcolor{gray!10}$a$          & CI Tau     & V* CI Tau       & K4IV   & 1      & Taurus       & $160.3_{-0.5}^{+0.5}$     & $1.4_{-0.3}^{+0.3}$       & 0.95  & 1.3--2.1   & 2021-12-09 & 32   &          1952 &              SV* SVS 1321 &           768\\
 $b$                           & CQ Tau     & V* CQ Tau       & F5IV   & 2      & Taurus       & $149.4_{-1.3}^{+1.3}$     & $6.3_{-0.3}^{+0.3}$       & 1.45  & 12.6--14.2 & 2021-01-01 & 16   &          2240 &            TYC 1865-648-1 &           448\\
\rowcolor{gray!10}$\cdots$     & CY Tau     & V* CY Tau       & M1.5   & 3      & Taurus       & $126.3_{-0.3}^{+0.3}$     & $0.48_{-0.05}^{+0.05}$    & 0.5   & 1.0--1.5   & 2021-12-09 & 32   &           544 &         J04170622+2802326 &           128\\
\rowcolor{gray!10}             &            &                 &        &        &              &                           &                           &       &            & 2021-12-27 & 32   &          3168 &         J04170622+2802326 &           768\\
 $c$                           & DL Tau     & V* DL Tau       & K7V    & 4      & Taurus       & $159.9_{-0.5}^{+0.5}$     & $0.81_{-0.13}^{+0.13}$    & 0.6   & 1.0--1.3   & 2021-12-04 & 32   &          3072 &           GSC 01833-00780 &           768\\
\rowcolor{gray!10}$d$          & DM Tau     & V* DM Tau       & M2V    & 4      & Taurus       & $144.0_{-0.5}^{+0.5}$     & $0.20_{-0.03}^{+0.03}$    & 0.6   & 5.0--8.0   & 2020-12-19 & 16   &          2256 &           GSC 01270-01088 &           448\\
 $\cdots$                      & DN Tau     & V* DN Tau       & M1:V   & 4      & Taurus       & $128.6_{-0.4}^{+0.4}$     & $0.81_{-0.06}^{+0.06}$    & 0.5   & 0.7--1.0   & 2021-11-24 & 32   &          2048 &          UCAC4 570-011400 &           384\\
                               &            &                 &        &        &              &                           &                           &       &            & 2021-12-10 & 32   &          3072 &          UCAC4 570-011400 &           768\\
\rowcolor{gray!10}$e$          & DS Tau     & V* DS Tau       & K4V    & 5      & Taurus       & $158.4_{-0.5}^{+0.5}$     & $1.03_{-0.16}^{+0.16}$    & 1.0   & 1.9--3.1   & 2021-12-29 & 32   &          3072 &          UCAC4 600-015051 &           768\\
 $f$                           & GM Aur     & V* GM Aur       & K3V    & 4      & Taurus       & $158.1_{-1.2}^{+1.2}$     & $1.4_{-0.18}^{+0.18}$     & 0.9   & 1.0--1.6   & 2021-01-20 & 16   &          2240 &         J04551015+3021333 &           448\\
\rowcolor{gray!10}$g$          & HD 31648   & HD  31648       & A5V    & 2      & Taurus       & $156.2_{-1.3}^{+1.3}$     & $19.9_{-0.7}^{+0.7}$      & 2.1   & 6.3--6.6   & 2021-12-10 & 16   &          3072 &                 HD 282758 &           768\\
 $h$                           & HD 34282   & V* V1366 Ori    & B9.5V  & 6      & Orion        & $309_{-2}^{+2}$           & $17.1_{-1.1}^{+1.1}$      & 2.0   & 9.7--14.3  & 2020-11-28 & 16   &          1120 &                BD-10 1143 &           192\\
                               &            &                 &        &        &              &                           &                           &       &            & 2020-12-24 & 16   &           512 &                BD-10 1143 &            64\\
                               &            &                 &        &        &              &                           &                           &       &            & 2020-12-27 & 16   &          2240 &                BD-10 1143 &           496\\
\rowcolor{gray!10}$i$          & HD 97048   & HD  97048       & A0V    & 7      & Cha          & $184.4_{-0.8}^{+0.8}$     & $47_{-7}^{+7}$            & 2.6   & 3.8--4.2   & 2021-01-28 & 16   &          2240 &                 CD-76 498 &           448\\
 $j$                           & HD 100453  & HD 100453       & A9V    & 8      & Lower Cen    & $103.8_{-0.2}^{+0.2}$     & $6.1_{-0.10}^{+0.10}$     & 1.6   & 10.0--16.0 & 2022-06-09 & 32   &          3200 &                 HD 100541 &           768\\
\rowcolor{gray!10}$k$          & HD 100546  & HD 100546       & A0V    & 9      & Lower Cen    & $108.1_{-0.4}^{+0.4}$     & $27.4_{-0.3}^{+0.3}$      & 2.2   & 5.0--7.0   & 2020-12-22 & 16   &          2240 &                 HD 101869 &           384\\
 $l$                           & HD 143006  & HD 143006       & G5IV   & 10     & Upper Sco    & $167.3_{-0.5}^{+0.5}$     & $3.6_{-0.3}^{+0.3}$       & 1.4   & 9.5--14.0  & 2021-06-30 & 16   &           704 &                BD-21 4234 &           320\\
                               &            &                 &        &        &              &                           &                           &       &            & 2021-07-22 & 16   &          2048 &                BD-21 4234 &           576\\
\rowcolor{gray!10}$m$          & HD 163296  & HD 163296       & A1V    & 2      & -            & $101.0_{-0.4}^{+0.4}$     & $15.9_{-0.3}^{+0.3}$      & 2.0   & 10.0--12.0 & 2021-04-06 & 16   &           384 &                 HD 163246 &           256\\
\rowcolor{gray!10}             &            &                 &        &        &              &                           &                           &       &            & 2021-06-03 & 16   &          2048 &                 HD 313493 &           576\\
\rowcolor{gray!10}             &            &                 &        &        &              &                           &                           &       &            & 2021-09-09 & 16   &          1536 &                 HD 313493 &           448\\
\rowcolor{gray!10}             &            &                 &        &        &              &                           &                           &       &            & 2021-09-26 & 16   &           512 &                 HD 313493 &           128\\
\rowcolor{gray!10}             &            &                 &        &        &              &                           &                           &       &            & 2022-06-11 & 64   &          3072 &                 HD 313493 &          1024\\
\rowcolor{gray!10}             &            &                 &        &        &              &                           &                           &       &            & 2022-07-07 & 64   &          3072 &                 HD 313493 &          1024\\
 $n$                           & HD 169142  & HD 169142       & F1V    & 9      & Upper Sco    & $114.9_{-0.4}^{+0.4}$     & $5.14_{-0.07}^{+0.07}$    & 1.6   & 11.2--23.1 & 2021-09-06 & 16   &          2048 &                 HD 169141 &           576\\
\rowcolor{gray!10}$o$          & IP Tau     & V* IP Tau       & M0:V   & 11     & Taurus       & $129.4_{-0.3}^{+0.3}$     & $0.38_{-0.03}^{+0.03}$    & 0.6   & 1.9--3.1   & 2021-12-09 & 32   &           512 &                     JH 33 &           128\\
\rowcolor{gray!10}             &            &                 &        &        &              &                           &                           &       &            & 2021-12-28 & 32   &          3072 &         J04284090+2655414 &           768\\
 $p$                           & IQ Tau     & V* IQ Tau       & M0.5   & 3      & Taurus       & $131.5_{-0.6}^{+0.6}$     & $0.70_{-0.11}^{+0.11}$    & 0.5   & 0.8--1.0   & 2022-01-01 & 32   &           960 &         J04284090+2655414 &           256\\
                               &            &                 &        &        &              &                           &                           &       &            & 2022-01-03 & 32   &           704 &         J04284090+2655414 &           128\\
                               &            &                 &        &        &              &                           &                           &       &            & 2022-01-06 & 32   &          3072 &         J04284090+2655414 &           768\\
\rowcolor{gray!10}$q$          & LkCa 15    & EM* LkCa   15   & K5:V   & 11     & Taurus       & $157.2_{-0.7}^{+0.7}$     & $1.17_{-0.11}^{+0.11}$    & 1.3   & 6.3--8.7   & 2020-11-27 & 16   &           768 &            TYC 1279-203-1 &           176\\
\rowcolor{gray!10}             &            &                 &        &        &              &                           &                           &       &            & 2020-12-08 & 16   &          2240 &            TYC 1279-203-1 &           448\\
 $r$                           & LkHa 330   & EM* LkHA  330   & F7     & 12     & Perseus      & $318_{-3}^{+3}$           & $20_{-6}^{+6}$            & 2.7   & 1.7--1.9   & 2020-12-08 & 16   &          2240 &         J03471855+3152187 &           448\\
\rowcolor{gray!10}$s$          & MWC 758    & HD  36112       & A8V    & 8      & Taurus       & $155.9_{-0.8}^{+0.8}$     & $10.3_{-0.3}^{+0.3}$      & 1.75  & 13.1--13.5 & 2020-12-19 & 16   &          2240 &                 HD 244395 &           448\\
\rowcolor{gray!10}             &            &                 &        &        &              &                           &                           &       &            & 2020-12-23 & 16   &          1024 &                 HD 244395 &           256\\
\rowcolor{gray!10}             &            &                 &        &        &              &                           &                           &       &            & 2020-12-26 & 16   &          2240 &                 HD 244395 &           448\\
 $t$                           & PDS 66     & CPD-68  1894    & K1V    & 1      & $\epsilon$ Cha & $97.89_{-0.12}^{+0.12}$   & $0.89_{-0.09}^{+0.09}$    & 1.2   & 4.9--7.2   & 2021-06-04 & 16   &          1536 &            TYC 9246-822-1 &           320\\
\rowcolor{gray!10}$u$          & PDS 201    & V* V351 Ori     & A7V    & 8      & Orion        & $326_{-3}^{+3}$           & $9_{-2}^{+2}$             & 1.7   & 9.9--13.5  & 2022-02-07 & 32   &          1024 &                 HD 290774 &           384\\
 $v$                           & SAO 206462 & CPD-36  6759    & F8V    & 13     & Upper Cen    & $135.0_{-0.4}^{+0.4}$     & $7.0_{-0.3}^{+0.3}$       & 1.55  & 9.6--10.0  & 2021-06-04 & 16   &          2048 &                 HD 135985 &           576\\
\rowcolor{gray!10}$\cdots$     & SR 20      & EM* SR   20     & G7     & 14     & $\rho$ Oph   & $138.1_{-0.6}^{+0.6}$     & $5.5_{-3.4}^{+3.4}$       & 1.9   & 3.1--4.0   & 2022-05-12 & 16   &           800 &              WMR2005 3-26 &           128\\
 $w$                           & SY Cha     & V* SY Cha       & K5V    & 15     & Cha          & $180.7_{-0.4}^{+0.4}$     & $0.72_{-0.10}^{+0.10}$    & 0.7   & 1.6--2.0   & 2021-01-02 & 16   &          2240 &         J11044460-7706240 &           448\\
\rowcolor{gray!10}$x$          & SZ Cha     & V* SZ Cha       & K0     & 16     & Cha          & $190.2_{-0.9}^{+0.9}$     & $2.6_{-0.2}^{+0.2}$       & 1.5   & 1.9--2.3   & 2020-12-29 & 16   &          1040 &              UCAC2 589393 &           256\\
\rowcolor{gray!10}             &            &                 &        &        &              &                           &                           &       &            & 2020-12-30 & 16   &          2240 &              UCAC2 589393 &           448\\
 $y$                           & V1094 Sco  & V* V1094 Sco    & K6     & 17     & Lupus        & $154.8_{-0.8}^{+0.8}$     & $0.64_{-0.07}^{+0.07}$    & 0.9   & 3.2--4.1   & 2021-09-10 & 16   &          1024 &           TYC 7855-1179-1 &           576\\
\rowcolor{gray!10}$z$          & V1247 Ori  & V* V1247 Ori    & F0V    & 8      & Orion        & $401_{-3}^{+3}$           & $16.3_{-0.8}^{+0.8}$      & 1.8   & 7.7--8.3   & 2020-12-20 & 16   &           512 &                 HD 290737 &            64\\
\rowcolor{gray!10}             &            &                 &        &        &              &                           &                           &       &            & 2020-12-22 & 16   &           512 &                 HD 290737 &           112\\
\rowcolor{gray!10}             &            &                 &        &        &              &                           &                           &       &            & 2020-12-24 & 16   &          2240 &                 HD 290737 &           448\\ \hline
\end{tabular}

\begin{flushleft}

{\tiny \textbf{Notes}: Column (1): Letter identifiers of the hosts in this paper, the $\cdots$ symbols are used for systems with no existing polarized observations in other bands or without confident detection in $K_s$-band for polarized color extraction. Column (2): Host name. Column (3): SIMBAD name in \citet{simbad}.  Column (4): Spectral type. Column (5): Spectral type reference in Column (4). Column (6): Region.  Column (7): Distance computed from \textit{Gaia} DR3 parallaxes \citep{GaiaDR3}. Column (8): Stellar luminosity. Column (9): Stellar mass. Column (10): System age. Column (11): UT date of observation. Column (12): Detector integration time (DIT) for both the host and the reference stars. The corresponding number of DIT values are NDIT = 1. Column (13): Total on-source exposure time for the host. Column (14): PSF reference star. Column (15): Total on-source exposure time for the PSF reference.
}

{\tiny \textbf{References}: References in Column (4) are 1: \citet{2006AA...460..695T}, 2: \citet{2001AA...378..116M}, 3: \citet{2012AA...538L...3R}, 4: \citet{1977ApJ...214..747H}, 5: \citet{1949ApJ...110..424J}, 6: \citet{1999MSS...C05....0H}, 7: \citet{1977PASP...89..347I}, 8: \citet{2003AJ....126.2971V}, 9: \citet{2017AJ....154...31G}, 10: \citet{2016MNRAS.461..794P}, 11: \citet{1986AJ.....91..575H}, 12: \citet{2014ApJ...786...97H}, 13: \citet{1995MNRAS.274..977C}, 14: \citet{2005AJ....130.1733W}, 15: \citet{2015AA...575A...4F}, 16: \citet{1980AJ.....85..444R}, 17: \citet{2017ApJ...847...31M}.
}
\end{flushleft}

\end{table*}

\subsection{Observations}
We observed 29 disks between 2020 November 27th and 2022 July 7th, using VLT/SPHERE in the star-hopping mode in $K_s$-band with the Infra-Red Dual-band Imager and Spectrograph \citep[IRDIS;][]{Dohlen08}. The data were obtained through six programs: \programESO{0103.C-0470}, \programESO{105.209E}, \programESO{105.20HV}, \programESO{106.21HJ}, and \programESO{108.22EE} (PI: M.~Benisty), and \programESO{105.20JB} (PI: M.~Keppler). 

By observing these systems in $K_s$-band, we can image the selected protoplanetary disk hosts in the longest wavelength offered by IRDIS.\footnote{\url{https://www.eso.org/sci/facilities/paranal/instruments/sphere/inst/filters.html}\label{fn-sphere-filters}} In this way, we can both reach a contrast regime that is more suitable in giant exoplanet imaging \citep[e.g.,][]{spiegel12, currie23pp7}, and simultaneously image circumstellar disks and complement existing IRDIS studies in shorter wavelengths.

The observations were conducted in the DPI mode with pupil tracking, so that we can simultaneously obtain both polarized light and total intensity observations for these systems.  In the star-hopping mode, we expect to obtain quasi-simultaneous capture of wavefront variations for an observation pair of a disk-hosting star (hereafter ``host'') and its corresponding well-chosen point spread function star (which does not host a disk or companion, hereafter ``reference''). With star-hopping, we can better capture the stellar speckles for a disk host using the reference  -- which has similar color, magnitude, and in close proximity of the corresponding host -- to better reveal circumstellar structures and exoplanets \citep{wahhaj21} than existing archival studies \citep[e.g.,][]{xie22}. To find the reference stars in our program, we used the \texttt{SearchCal} tool of the JMMC.\footnote{\url{http://www.jmmc.fr/searchcal}\label{fn-jmmc}} See Table~\ref{tab:prop} for the reference stars, which have identical observational setup as their corresponding host stars, used in our observations.

\subsection{Data reduction}
We retrieved the raw data in \texttt{fits} format \citep{pence10} for our programs from the ESO archive for SPHERE.\footnote{\url{http://archive.eso.org/wdb/wdb/eso/sphere/form}} To reduce the IRDIS data, we proceeded as detailed in the following sub-sections: we first used the \texttt{IRDAP} \citep{irdap1, irdap2} package\footnote{\url{https://irdap.readthedocs.io/en/latest/}}  for polarimetric differential imaging, which includes both pre-processing of the raw data in Sect.~\ref{sec-preprocess} and post-processing in polarized light, see Sect.~\ref{sec-pdi}. We then performed reference differential imaging in total intensity light using the \texttt{IRDAP} output files, see Sect.~\ref{sec-rdi}. The final data products in \texttt{fits} format are publicly available at the CDS via anonymous ftp to \url{cdsarc.cds.unistra.fr} (\url{130.79.128.5}) or via \url{https://cdsarc.cds.unistra.fr/viz-bin/cat/J/A+A/} (link to be specified upon manuscript publication). 

\subsubsection{Preprocessing}\label{sec-preprocess}
In the star-hopping observation of a host-reference pair, the host has dedicated sky background observations for empirical $K_s$-band background removal. To remove the stellar signals and speckles in host images, reference stars serve as empirical point spread functions (PSFs; they are occulted by the coronagraph here), yet they do not have dedicated sky backgrounds. To enable sky background removal for a reference star, we used the sky frames of its corresponding host.

We customized the preprocessing procedure in \texttt{IRDAP} to reduce the star-hopping datasets. In addition to sky background removal for the reference, we recalculated the parallactic angle for the target exposures from the \texttt{PynPoint}\footnote{\url{https://pynpoint.readthedocs.io/en/latest/}} \citep{pynpoint} pipeline for SPHERE. We also rescaled the preprocessed rectangular IRDIS pixels to square pixels by streching the pixels by $1.006$ along the vertical direction of the detector \citep[e.g.,][]{schmid18}. The pre-processed images from \texttt{IRDAP} are $1024\times1024$~pixel, where 1 IRDIS pixel is 12.25 mas \citep{maire16}, with the stars located at the centers of the images.

\begin{figure*}[htb!]
\centering
	\includegraphics[width=\textwidth]{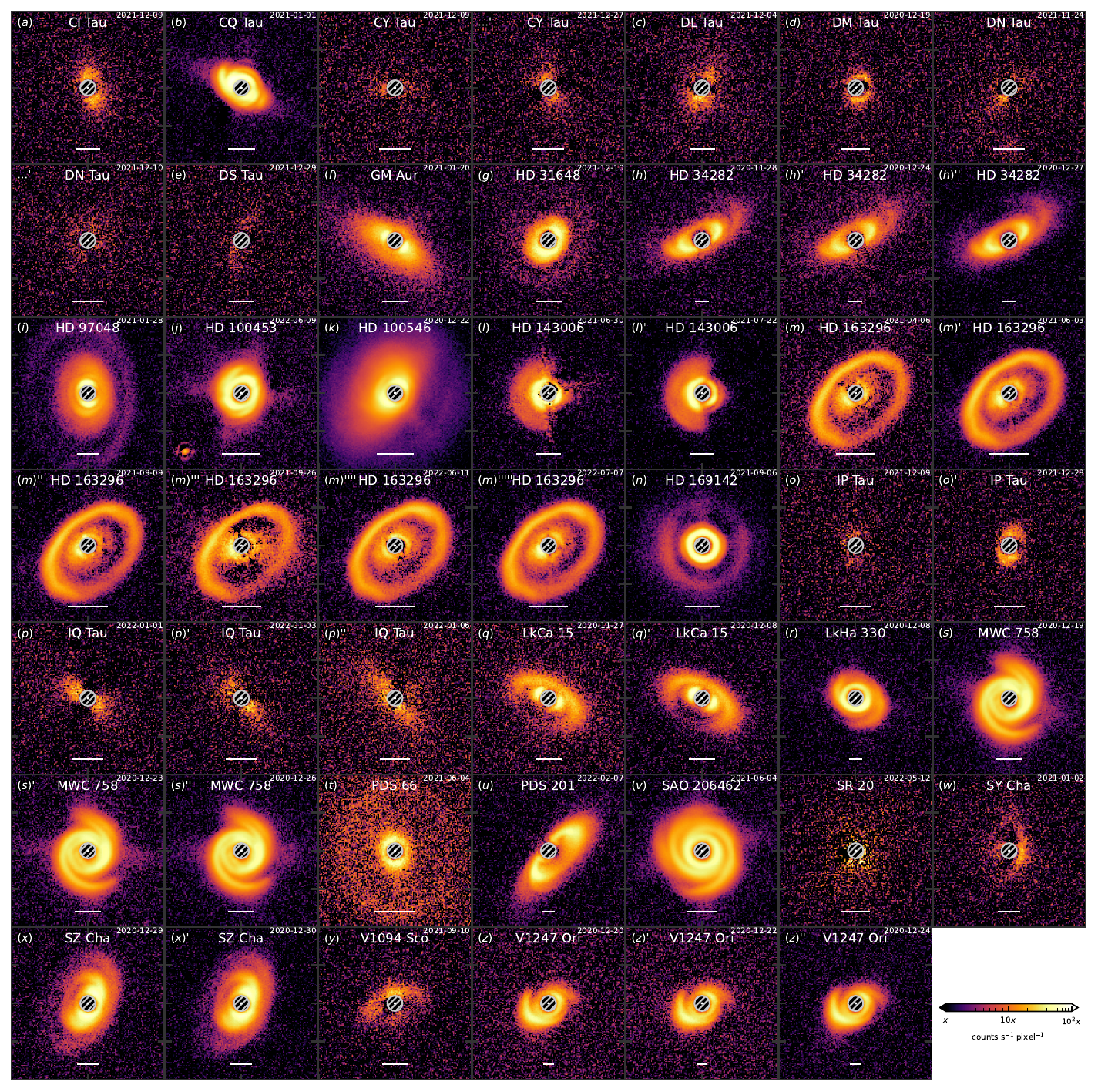}
    \caption{$K_s$-band \Qphi\ maps with dimensions of $2\arcsec{\times}2\arcsec$ with different color bars in log scale. The letter identifiers are from Table~\ref{tab:prop}. The rulers correspond to $50$~au. The regions interior to $0\farcs1$ are not accessible with coronagraph usage.\\ (The data used to create this figure are available.)}
    \label{fig-qphi}
\end{figure*}

\begin{figure*}[htb!]
\centering
	\includegraphics[width=\textwidth]{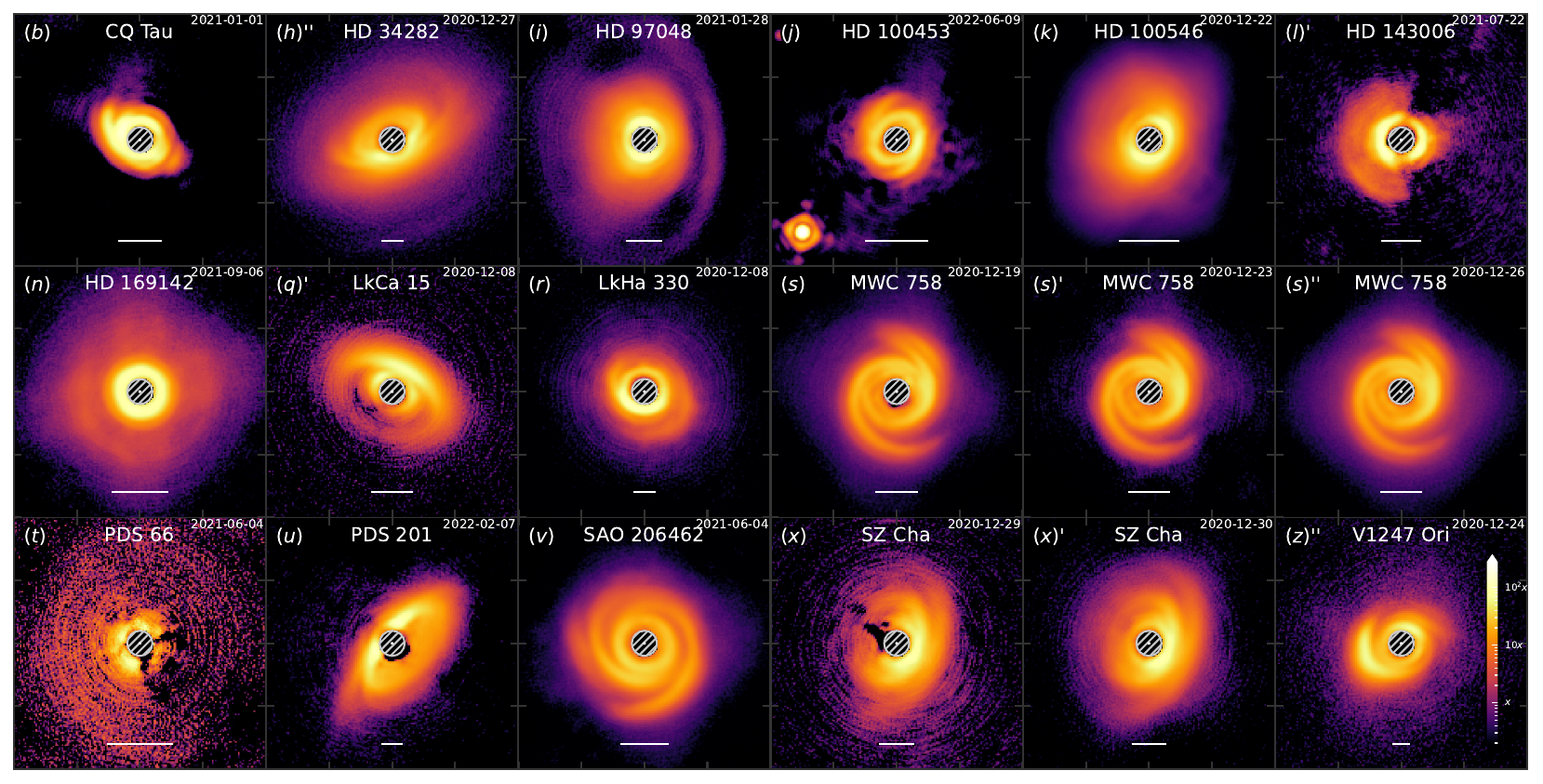}
    \caption{$K_s$-band \Itot\ maps with dimensions of $2\arcsec{\times}2\arcsec$ with different color bars in log scale, for disks identifiable with DI-sNMF star-hopping RDI data reduction.  Due to varying observation conditions, a well-chosen reference star may not lead to optimum reduction results, e.g., SZ~Cha in panels ($x$) and ($x$)'. The data including those with non-identifiable total intensity detections are available online in \texttt{\texttt{fits}} format.\\ (The data used to create this figure are available.)}
    \label{fig-tot}
\end{figure*}

\subsubsection{Polarimetric differential imaging (PDI): \Qphi\ images}\label{sec-pdi}

For polarized data, we performed PDI data reduction using \texttt{IRDAP}. We ran \texttt{IRDAP} with the default set of parameters to perform PDI reduction. We use the star-polarization-subtracted \Qphi\ data for further analysis, see Fig.~\ref{fig-qphi}. We also reduced available archival observations with DPI in other IRDIS bands (i.e., $Y$, $J$, or $H$), see Appendix~\ref{app-pol-aux} and Table~\ref{tab:archive}, to allow for a comparison with the $K_s$-band data from this study.

\subsubsection{Reference diffraction imaging (RDI): \Itot\ images}\label{sec-rdi}
For total intensity data that can be generated using the polarized observations, we performed RDI data reduction to obtain total intensity (\Itot) images of the systems using the data imputation technique described in \citet{ren20di}. For this purpose, we added the \texttt{IRDAP} preprocessed images from the left and the right IRDIS channels, then used the central $350\times350$~pixel for RDI post-processing. 

For the reference cube, we obtained their exposure features using non-negative matrix factorization \citep[NMF:][]{ren18nmf} with 5 sequentially constructed NMF components (i.e., sNMF). In this work, we experimented using 10 or more sNMF components for data interpretation, yet we did not observe clear difference or improvement in the quality of RDI outputs, we thus adopted 5 sNMF components for the rest of the study for computational efficiency. We note that this is due to the high speckle similarity of star-hopping observations, and thus for non-star-hopping observations (e.g., archival data analysis), more sNMF components should be used \citep[e.g.,][]{xie23, Krishnanth23}.

To obtain the \Itot\ disk images from the disk host cubes, we imputed stellar signals in a circular region that is within a radius of 80 pixel from the center using the sNMF components. Specifically, we used an annular region that is between 80 pixels and 175 pixels from the stars to model the entire field of view using the NMF data imputation (DI-sNMF) approach described in \citet{ren20di}. We chose the annular region since it both covers the control ring of the adaptive optics (AO) system for SPHERE in $K_s$-band (the regions can change in different wavelengths), which contains sufficient information to infer the light distribution across the entire field of interest, and does not contain disk signals that generally lie within $1\arcsec$ (i.e., ${\sim}82$ IRDIS pixels). We then removed the DI-sNMF models for each image in the disk host cube, derotated them to north-up and east-left using previously calculated parallactic angles in Sect.~\ref{sec-preprocess}, and adopted the element-wise median as the final disk image in total intensity. We present the total intensity images using RDI from DI-sNMF in Fig.~\ref{fig-tot}.

\subsubsection{Polarization fraction and color}
With both polarized light and total intensity images, we computed the linear polarization fraction maps. We divided the linearly polarized \Qphi\ data with PDI from \texttt{IRDAP} by the \Itot\ data with RDI from DI-sNMF. With relatively negligible uncertainty in the \Qphi\ data from \texttt{IRDAP}, the corresponding uncertainty for the polarization fraction maps are propagated using the element-wise standard deviation map of the \Itot\ results.

To obtain the relative reflectance of the disks, we divided the \Qphi\ images by the \texttt{IRDAP} measurement of the star. We removed shot noises in the \Qphi\ data with two-dimensional Gaussian profile smoothing with standard deviations of $\sigma=2$~pixel \citep[e.g.,][]{Olofsson18}. For observations in non-$K_s$ bands, we additionally convolved the data to reach same spatial resolution as $K_s$-band, based on Rayleigh diffraction limits of $1.22\lambda/D$ where $\lambda$ is the central wavelength\textsuperscript{\ref{fn-sphere-filters}} and $D$ is the telescope pupil size of $8.0$~m. To obtain disk colors in \Qphi, we converted the relative reflectance to magnitudes, then subtracted shorter wavelength magnitudes from longer wavelength ones.

\section{Disk imaging}\label{sec:analysis}

We gather the star-polarization-subtracted \Qphi\ results from \texttt{IRDAP}, and present the images in Fig.~\ref{fig-qphi}. To our knowledge, the maps of 10 objects were never published before: these are CY~Tau, DL~Tau, DM~Tau, DN~Tau, DS~Tau, IP~Tau, and IQ~Tau, SR~20, SY~Cha and SZ~Cha. For systems with high-quality in total intensity \Itot\ detections from RDI data reduction using DI-sNMF, we present the images in Fig.~\ref{fig-tot}. 

\subsection{PDI \Qphi\ maps}
We obtained disk detection for nearly all of the targets in Fig.~\ref{fig-qphi}. For faint disks such as those of CI~Tau, CY~Tau, DL~Tau, DN~Tau, DS~Tau, IP~Tau, IQ~Tau, and SR~20, some of their observations data did not produce high-quality detections (e.g., IP~Tau) or even approaching non-detection (e.g., SR~20). In comparison with existing PDI data of the systems at shorter wavelengths (i.e., $Y$-, $J$-, or $H$-band), the \Qphi\ maps in $K_s$-band in Fig.~\ref{fig-qphi} do not have significant morphological variations from them. The $K_s$-band data are less resolved due to an increase in observation wavelength. In the gallery, it is apparent that the M-star companion HD~100453B is polarized in Fig.~\ref{fig-qphi}(j), and the polarization is also detected in \texttt{IRDAP}-reduced archival data in $J$- and $H$-band. A polarized HD~100453B indicates that it hosts dust, similarly to CS~Cha~b \citep{Ginski2018}.

\Qphi\ signals are expected to primarily trace single scattering events on the surfaces of optically thick disks. The signals are expected to be positive when the polarization vectors are penpendidular to the direction of the incident light on the scatters \citep[e.g.,][]{Monnier2019}. Nevertheless, multiple scattering naturally occurs in protoplanetary disks that are optically thick, reducing the polarization fraction of disks \citep[e.g.,][Fig.~4 therein]{Tazaki2019}. In observations, multiple scattering signals can be revealed in the \Uphi\ images \citep[e.g.,][]{Canovas2015, Monnier2019}, which traces the light that are ${\pm}45^\circ$ from the incident light. In addition, due to the finite angular resolution with VLT/SPHERE, \texttt{IRDAP} \Qphi\ maps for IRDIS in our study could be lower limits of the actual \Qphi\ light (i.e., $\hat{\mathcal{Q}}_\phi$ in \citealp{Ma2023}) due to convolution effects. Observationally, studies including \citet{Engler2023} demonstrated that the \Uphi\ maps, which in principal should not contain signals for single-scattering systems such as the HD~114082 debris disk, does contain spurious signals in SPHERE/IRDIS in $H$-band (central wavelength: ${\sim}1.625~\mu$m) but not SPHERE/ZIMPOL in $I\_PRIM$ band (${\sim}0.790~\mu$m), with the latter having higher angular resolution. For the $K_s$-band (${\sim}2.182~\mu$m) observations in this study, we thus expect the \Uphi\ maps contain more leakage from \Qphi\ data due to finite angular resolution. We observe that the \Uphi\ signals in Fig.~\ref{fig-uphi} are ${\lesssim}5\%$ of the \Qphi\ signals in absolute values, and thus the \Qphi\ images in this work are not severely impacted (see, e.g., \citealp{Canovas2015}, for non-negligible impacts in their simulation). To further reduce the limitation on \Uphi\ signal leakage, which are albeit of minor contribution in absolute values here, forward modeling of the convolution effects is necessary \citep[e.g.,][]{Engler2018, Tschudi2021, Ma2023}, and such a solution \citep[e.g.,][Section 4.2.1 therein]{Ma2023} is beyond the scope of this study.

\subsection{RDI \Itot\ maps and their fidelity}\label{sec-tot-disc}

Using the SPHERE control ring at a region of $80$--$110$~pixel ($0\farcs98$--$1\farcs35$ from the center) and the region exterior to it in $K_s$-band (i.e., $80$--$175$~pixel), we were able to recover disks in RDI total intensity in Fig.~\ref{fig-tot} for 15 systems using DI-sNMF. While these disks were detected in total intensity, the rest of the disks in our sample were detected with high fidelity primarily only in polarized light (e.g., CI~Tau, HD~163296). 

To obtain the RDI images, we only used the signals outside the inner edge of the IRDIS control ring of a host (i.e., $80$--$175$~pixel) to infer the speckles interior to the control ring. In other words, we used ${\sim}80\%$ of the image to infer the entire image, yet thus such an imputation might fall into the regime of under-fitting. In fact, we also used using only the control ring in $K_s$-band (i.e., $80$--$110$~pixel) for DI-sNMF reduction, and the results do not have significant change: this further supports the importance of the control ring in inferring the PSF signals interior to it using data imputation. In comparison, \citet{ren20di} showed that the region used for imputation has a $2^{\rm nd}$-order influence on the recovery quality of speckle signals. Therefore, with ${\sim}20$--$50\%$ of the regions being masked out in the study here, we would have expected a ${\sim}25\%$ change according to \citet{ren20di}, which should have resulted into inferior qualify for the imputed stellar signals and thus recovered disks. To investigate the mathematical reason for the high-quality images in Fig.~\ref{fig-tot}, we advanced the mathematical investigation by presenting a corresponding derivation for ideal imputation cases (i.e., when the \citealp{wahhaj21} requirements for reference stars are fully satisfied) in Appendix~\ref{app-di-ideal}. In fact, when the individual matrix elements in the matrices are weighted equally in DI-sNMF, the contribution of ``missing data'' can only have a theoretical $4^{\rm th}$-order impact. 

With the new derivation in Appendix~\ref{app-di-ideal} showing an expected $4^{\rm th}$-order deviation from the missing data, we now can further establish the mathematical background for the authenticity of DI-sNMF reduction, especially when data quality (i.e., speckle stability, speckle resemblance across disk host and reference stars) is ideal. Based on the quality of disk recovery in Fig.~\ref{fig-tot}, we now expect that the masked out region have a ${\sim}6\%$ difference even when ${\sim}50\%$ of the regions are masked out: this supports the observed high-fidelity morphological similarity between the PDI and RDI results. Moving forward, this indicates that for extended structures that overlap with the control ring, or for disk observations in shorter wavelengths where the control ring is angularly closer-in than the $K_s$-band data here, we do not have to use the entire control ring to recover the extended structures. We leave such an investigation to future work.

\subsection{RDI \Itot\ detectability with stellar parameters}\label{sec-tot-detect}

\begin{figure*}[htb!] 	
\centering
\includegraphics[width=\textwidth]{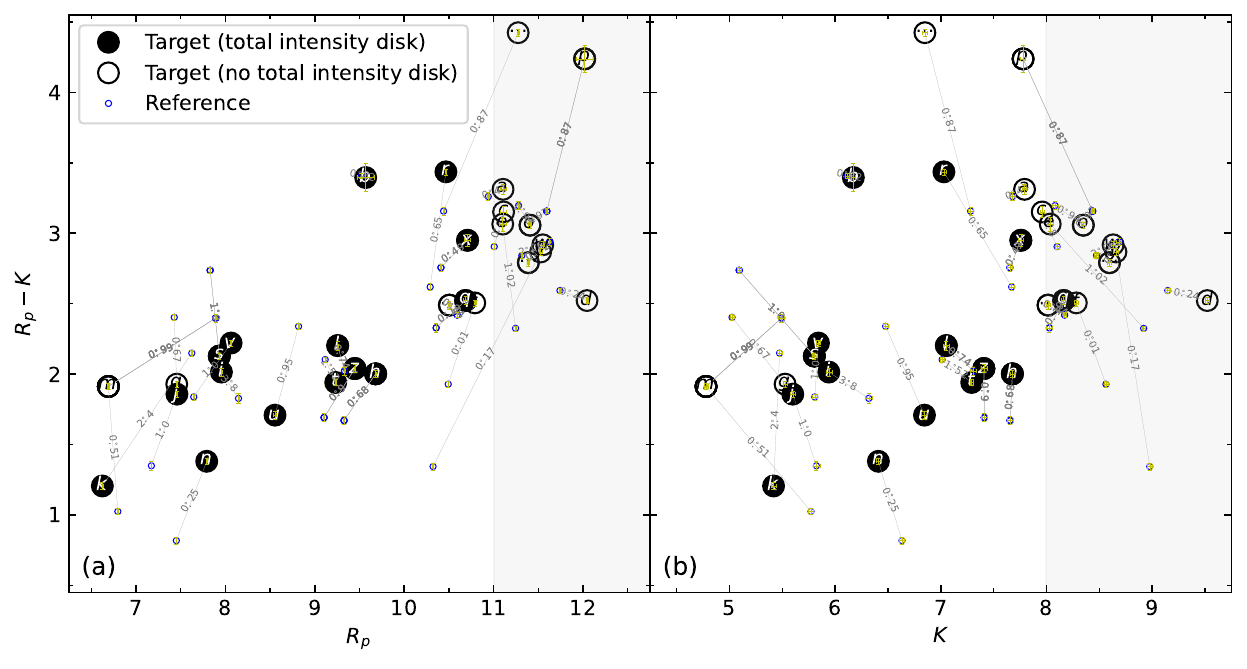}
    \caption{Disks with confident recovery in $K_s$-band via NMF data imputation, as a function of (a) \textit{Gaia}~DR3 $R_p$ magnitude or (b) 2MASS $K_s$ magnitude with the $R_p-K$ color. Each connected pair is a host-reference pair, with their on-sky angular separation from \textit{Gaia}~DR3 in degrees. Notes: (1) Certain systems with marginal detections in NMF data imputation are marked as non-detection (e.g., HD~163296 or $m$). (2) The size of reference star symbols reflects typical uncertainties in color-magnitude measurements, zoom in the figure for actual error bars colored yellow for all systems.}
    \label{fig-tot-detectability}
\end{figure*}

We show in Fig.~\ref{fig-tot-detectability} the detectability of disks in total intensity (the confirmed detections are from Fig.~\ref{fig-tot}), as a function of their stellar $R_p-K$ color and $R_p$-band (or $K$-band) magnitude. The $R_p$-band and $K$-band magnitudes are from \textit{Gaia}~DR3 \citep{GaiaDR3} and 2MASS \citep{2mass}, respectively.

Beyond an empirical threshold of $R_p\gtrsim11$ or $K\gtrsim8$, the disks are not detected in total intensity with DI-sNMF even when they are detected in \Qphi. Given that DI-sNMF depends on the adaptive optics' control ring signals for data reduction in Sect.~\ref{sec-tot-disc}, this illustrates that the importance of the  adaptive optics performance \citep[e.g.,][]{Jones2022} in producing the control ring for DI-sNMF data reduction.

\begin{figure*}[htb!] 	
\centering
\includegraphics[width=0.9\textwidth]{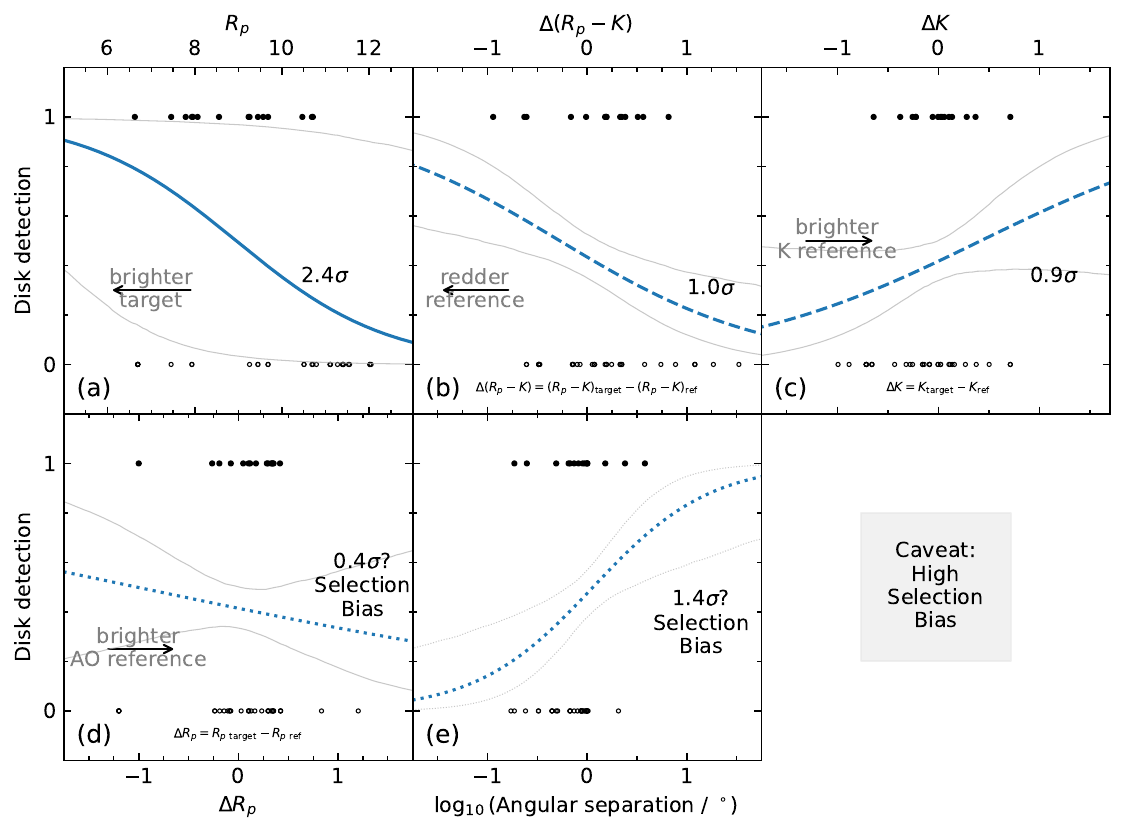}
    \caption{Individual logistic regression on RDI detectability of disks in DI-sNMF. The solid dots are for systems in Fig.~\ref{fig-tot}. (a) Brighter host stars can likely yield disk detection. (b) Redder $R_p-K$ band color potentially contribute to disk detection. (c) Brighter reference stars in $K_s$-band might tentatively contribute to disk detection. (d) Reference stars with fainter $R_p$ magnitude of target stars might tentatively contribute to disk detection, however that is likely biased by the non-detections with $R_p {\gtrsim} 11$, see Fig.~\ref{fig-tot-detectability}(a). (e) Although reference at farther separation might yield better disk detection, this stems from a selection bias when close-in references are not available. Notes: (1) A joint multinomial logistic regression only preserves the trend for (a). (2) The trends here are subject to change due to the highly selected samples in this study. (3) From solid to dashed lines, the statistical significance decreases; dotted lines are biased trends that are not trustworthy.}
    \label{fig-tot-detectability-logit}
\end{figure*}

\subsubsection{Logistic regression detectability}

To quantify disk detectability using other system parameters, we used \texttt{R} \citep{R} to perform logistic regression. We explored independent regression in Fig.~\ref{fig-tot-detectability-logit} using the host star magnitude, host-reference color difference, host-reference magnitude difference in $R_p$- and $K_s$-band, or angular separation between the target and the reference star. When we fit these parameters independently, we observe that brighter stars can yield better detection, see Fig.~\ref{fig-tot-detectability-logit}(a). While redder reference and brighter $K_s$-band reference could likely yield better detection, see Fig.~\ref{fig-tot-detectability-logit}(b) and (c) respectively, the relations are marginal.  

We originally observed that fainter reference in $R_p$-band or angularly closer-in references might negatively impact detection, see Fig.~\ref{fig-tot-detectability-logit}(d) and (e) respectively. However, we argue that they result from selection bias. On the one hand, most of the non-detections with $R_p \gtrsim11$ have brighter references, which bias the fit in Fig.~\ref{fig-tot-detectability-logit}(d). On the other hand, specifically, the HD~97048 disk is detected in Fig.~\ref{fig-tot}(i) while the reference is $3\fdg80$ from it in Fig.~\ref{fig-tot-detectability} (that reference star is chosen since there are no other good nearby references for HD~97048). As a result, to select reference stars for star-hopping observations, we do not recommend selecting fainter $R_p$-band references or angularly distant references. Instead, we recommend to focus on other parameters in Fig.~\ref{fig-tot-detectability-logit}(a)--(c): brighter host stars, redder references, and slightly brighter references in observational wavelengths.

When we jointly fit the disk detectability with DI-sNMF in total intensity with these parameters, only the relationship on target star magnitude persist, and that trend does not change when we use the \textit{Gaia} $G$-band\footnote{We adopt \textit{Gaia} $R_p$-band here, since \citet{wahhaj21} referenced $R$-band for star-hopping observations.} or 2MASS $K_s$-band magnitudes. After all, the reference stars were already selected based on their magnitude and color match, as well as on-sky proximity, the logistic regression here could suffer from severe selection bias that is less severe were the references chosen randomly. However, a random selection of reference stars is dissuaded in star-hopping observations due to observation efficiency.

Due to the high selection bias of the sample in this work, we did not explore the detectability of disks as a function of disk property such as mass and inclination. To perform this study, we first need a proper removal of dominating effects such as host star brightness in Fig.~\ref{fig-tot-detectability-logit}(a). However, due to the limited RDI disk detections here with DI-sNMF, we do not have enough targets for the exploration on disk properties (see Sect.~4.3 of \citealp{Ren23} for a similar discussion).

\subsubsection{Implications for star-hopping reference selection}

Assuming the relationships in Fig.~\ref{fig-tot-detectability-logit}(a)--(c) are trustworthy, we explain the RDI disk detectability as follows. First, brighter disk-hosting stars in apparent light in Fig.~\ref{fig-tot-detectability-logit}(a) can offer more light for scattering in disk particles, increasing disk detectability. Second, the AO system of SPHERE operates in visible wavelength that overlaps with the $R_p$- and $G$-band of \textit{Gaia} \citep[e.g.,][ Fig.~5 therein]{Jones2022}, and thus AO operations could potentially offer similar deformable mirror corrections in a star-hopping observation, especially when the reference stars is of similar magnitude as the target star (or slightly brighter to ensure comparable AO performance). Third, assuming there is a pair of host and reference stars with similar magnitudes in AO operation wavelengths, then the redder the reference is (i.e., brighter in $K_s$-band), the higher signal-to-noise ratio for the reference exposures, and thus the better DI-sNMF data reduction quality. In comparison, for broader-band filters such as the \textit{Hubble Space Telescope}'s Space Telescope Imaging Spectrograph, which operates at ${\sim}0.2~\mu$m -- ${\sim}1.2~\mu$m, using color-matching and slightly brighter references can also yield better host signal recovery \citep{Debes2019}. Following these arguments, the detection of HD~163296 with low fidelity can be explained in Fig.~\ref{fig-tot-detectability}: one reference in $R_p$-band is $1.2$ magnitude fainter than HD~163296 and thus yielded different AO operational status, the other reference (which has similar magnitude and redder) did not produce high-fidelity results, since it was resolved by IRDIS as a binary system during the observation. 

There is no clear evidence that on-sky proximity would enhance RDI disk detection. Therefore, star-hopping users should attribute a low priority to it in their reference star selection, and instead focus on the above-mentioned parameters. Nonetheless, it is not clear if planet detection capability is impacted by on-sky proximity to the host in star-hopping, especially when there is no circumstellar disks.

For systems with control rings with high signal-to-noise ratios, the fact that the two IRDIS channels pass through different optical paths may yield slight difference in data reduction quality. A potential increase of the disk quality in this work is to perform DI-sNMF reduction for the two channels separately. However, this approach is beyond the current scope of this study, since the purpose of this section is to demonstrate the usage of the control ring in DI-sNMF data reduction. For this purpose, we added the data from the two IRDIS channels, to obtain a factor of ${\sim}\sqrt{2}$ increase in the signal-to-noise ratio for the pixels containing control ring signals.

\section{Companion search}\label{sec:contrast}
From RDI results, we did not identify companions except for HD~100453. HD~100453 hosts two prominent spirals and has an M star companion \citep[i.e., HD~100453B, e.g.,][]{wagner15, benisty17}, and the companion has a $K_s$-band polarization fraction of $1.0_{-1.0}^{+1.5}\%$, which is measured here using the PDI total polarized intensity and RDI total intensity results. The polarized HD~100453B suggests that there exists circumsecondary disk. See \citet{vanHolstein2021} for a study on polarized companions using IRDIS.

In this section, we obtain the detection limits for our star-hopping RDI datasets. We also compare our $K_s$-band results with previous claims, and discuss potential reasons of their non-detection in $K_s$-band here.

\subsection{Detection limits}

To quantify the detection capability of point sources, existing surveys \citep{Nielsen2019, Vigan21} adopted primarily the principal-component-analysis-based algorithms \citep[e.g., KLIP:][]{amara12, soummer12}. However, the results from these methods are prone to contamination, especially when disk signals exist, which further prevent the detection of companions from a combination of algorithm choice (overfitting or over-subtraction) and the ADI observational strategy (self-subtraction). As a result, aggressively post-processed disk signals can resemble point sources \citep[e.g.,][]{Rameau2017, Currie2019}, and extensive optimization is needed to increase the significance level of detections \citep[e.g.,][]{adams2023}.

With the RDI results from DI-sNMF on star-hopping data, we can now better preserve disk signals to avoid them being regarded as of planetary origin. For the RDI data from DI-sNMF here, we do not see significant improvement when more than 5 sNMF components are used, and thus we use them to calculate the detection limits of companions as a function of angular separation from the star.

\begin{figure*}[t!] 	
\includegraphics[width=0.5\textwidth]{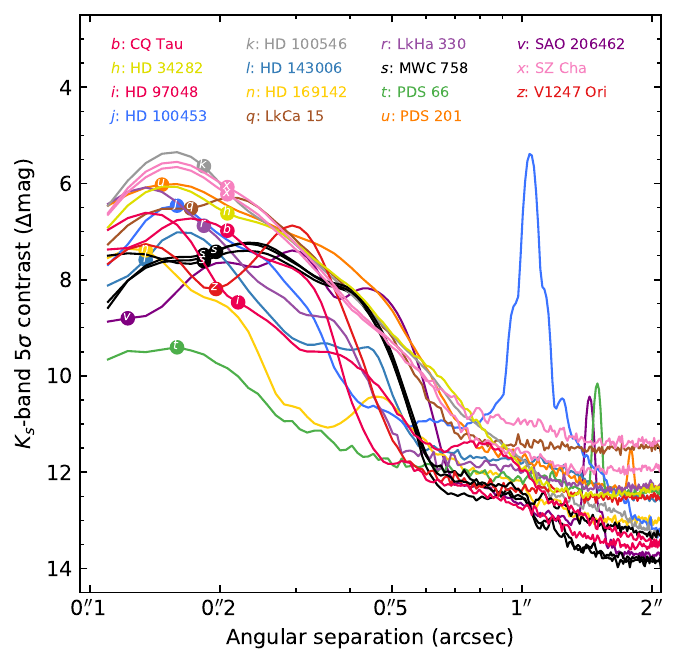}
\includegraphics[width=0.515\textwidth]{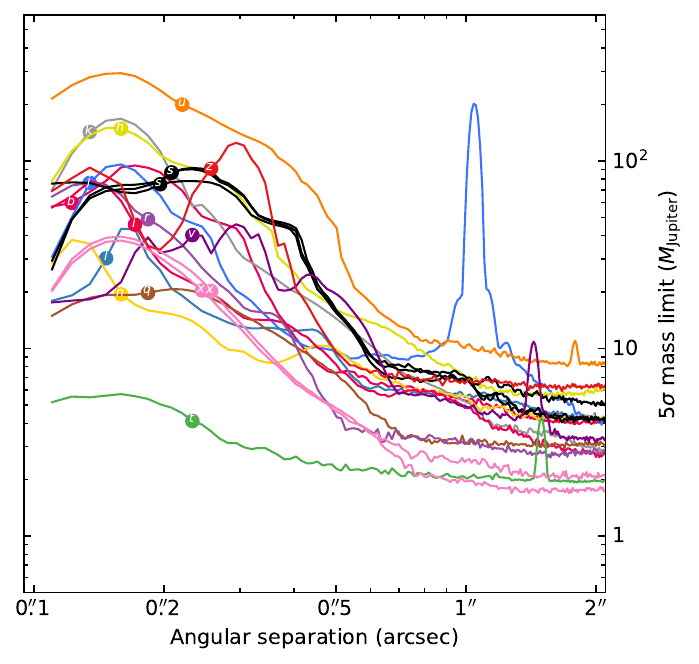}

    \caption{RDI contrast curves (left) and AMES-Cond mass limits (right) of DI-sNMF reductions for systems in Fig.~\ref{fig-tot}, see Sect.~\ref{sec:contrast} for the detailed calculation procedure. Interior to ${\sim}0\farcs2$, the measured contrast is prone to transmission reduction near the coronagraphic edge. The contrast peak at ${\sim}1\arcsec$ for $j$ (HD~100453) is the binary companion. The contrast curve for $t$ (PDS~66) is the deepest due to the faintness of the disk.\\ (The data used to create this figure are available.)}
    \label{fig-contrast}
\end{figure*}

\subsubsection{Contrast calculation}
Using the preprocessed and postprocessed data, we calculated the RDI contrast curves from DI-sNMF for the disks with high-quality detection from Fig.~\ref{fig-tot}. Using the RDI images, for one specific angular separation from the star, we calculated the radial profiles for the median and standard deviation within a 3 pixel annulus. We then rescaled the standard deviation profile by taking into account the small sample statistics in \citet{Mawet2014}. Given that these radial profiles are calculated in detector units, we obtained the ratio needed for contrast conversion using stellar counts as follows: using the preprocessed output from \texttt{IRDAP}, we obtain the peak-to-total ratio between the peak and the total stellar photometry from the \texttt{fits} files in the \texttt{calibration/flux} folder (i.e., the flux files and regions used by \texttt{IRDAP} for stellar flux measurement and background removal). We then converted the measured radial profiles to a ratio with the star by dividing them by the peak-to-total ratio.

Noticing the existence of disks, we present in Fig.~\ref{fig-contrast} the $5\sigma$ contrast curves by multiplying the rescaled standard deviation profile in units of ratio by $5$. For one $5\sigma$ contrast curve, we did not add its corresponding median profile which reflects the disk signal to it, since any point source with a brightness of $5$ times the rescaled standard deviation will be super imposed onto the disk, thus reaching a $5\sigma$ contrast.  It should be noted that the RDI results from DI-sNMF do not have a mean of zero for each reduction image as KLIP. Instead, the RDI contrast with DI-sNMF is calculated only using the standard deviation due to the existence of disks. To ensure the reliability of the DI-sNMF contrast, we compared our results with the TLOCI \citep{tloci} ones from the High Contrast Data Centre,\footnote{\url{https://sphere.osug.fr/spip.php?rubrique16&lang=en}} and did not observe significant differences.

\subsubsection{Contrast results}

For RDI with DI-sNMF, we reached for $K_s$-band a $5\sigma$ contrast of $\Delta{\rm mag} \approx 6$--$10$ at ${\sim}0\farcs2$, and $\Delta{\rm mag}\approx11$--$13$ at ${\sim}1\arcsec$. In comparison with \citet{wahhaj21}, where the detection limits are $\Delta{\rm mag} \approx 12$ and ${\sim}14$ correspondingly in $K_1$- or $K_2$-band with SPHERE/IRDIS,  the detection limits here are ${\sim}2$~mag or more brighter. We note that at least two factors have resulted into the difference. First, while the wavelength coverage of the two studies are similar\textsuperscript{\ref{fn-sphere-filters}}, the filters used in \citet{wahhaj21} have narrower wavelength coverage ($K_1$- and $K_2$-band) than the $K_s$-band here. With broader wavelength coverage, stellar speckles on detectors are more extended, thus limiting companion detection in reaching lower contrasts \citep[e.g.,][]{Groff2016, Desai2022}. Nevertheless, this does not suggest that narrower bands can provide better detections, since a detection is a trade off between the contrast and the planetary luminosity integrated in a band. Second, the existence of bright disks does limit companion detection \citep[e.g.,][]{Quiroz2022}, and the asymmetric distribution of disk signals additionally increases the standard deviation in a given annulus for contrast calculation. In fact, PDS~66 offers the best detection limit in Fig.~\ref{fig-contrast}, with $\Delta{\rm mag}\approx10$ at ${\sim}0\farcs2$ which is closer to the \citet{wahhaj21} values. This is because that the PDS~66 disk is fainter and more symmetric than other detected disks in total intensity, especially in comparison with other systems with similar $R_p$- and $K$-band stellar magnitudes (e.g., $h$, $l$, and $z$ in Figs.~\ref{fig-tot} and \ref{fig-tot-detectability}), and thus the impact of disk is smaller than other systems.

To compare the $K_s$-band data here with other observations, we used AMES-Cond models \citep{Allard2012} to convert the contrast curves to $5\sigma$ upper limits of mass following \citet{wallack23}. We used the mean and standard deviation of the two age estimates in Table~\ref{tab:prop} to calculate the mass limit for companions in Fig.~\ref{fig-contrast}. We note that we did not account for the extinction from the circumstellar disk, and thus the sensitivity can decrease with the existence of extinction \citep[e.g.,][]{Cugno2023}. However, the determination of system ages, together with the evolutionary models and their assumptions \citep[e.g.,][]{Allard2012, Baraffe2003}, could significantly change the detection limits \citep[e.g.,][]{AsensioTorres2021, wallack23}. We note that the limits that we derived are consistent with the ranges of estimates of previous studies \citep[e.g.,][see Fig.~7 therein]{AsensioTorres2021}. Nevertheless, the star-hopping mode should be preferred for future planet-hunting efforts, since it does not have a requirement on sky rotation for ADI data reduction.

We note that the star-hopping observations in this study were executed in the pupil-tracking mode of  SPHERE/IRDIS, for which both pupil-tracking and field-tracking modes are available \citep[e.g.,][]{maire21}. In pupil-tracking, the diffraction spikes that are evident in field-tracking are suppressed. In field-tracking, although we should obtain a sufficient sky rotation for ADI data reduction, the rotation of the diffraction spikes in an observation sequence would limit the data reduction quality in reducing star-hopping observations with RDI, since the diffraction spikes from a host image cannot be removed using a rotated set of diffraction spikes from a reference image. For RDI data reduction, we thus only recommend star-hopping under pupil-tracking.

\subsection{Comparison with existing claims}
Several targets in this study were reported to have exoplanet candidates or claims from high-contrast imaging, e.g., HD~100546 \citep{Quanz2015}, HD~169142 \citep{Hammond2023}, LkCa~15 \citep{Kraus2012, Sallum2015}, and MWC~758 \citep{Reggiani2018, Wagner2019, Wagner2023}. However, some of them were later identified to be likely disk signals (e.g., HD~100546: \citealp{Rameau2017}, LkCa~15: \citealp{Currie2019}) or non-confirmation \citep[e.g., MWC~758:][]{Boccaletti2021}. For the planets that are embedded in disks \citep[e.g., PDS~70~c;][]{Haffert2019}, proper separation of the signal between the planet and the disk is needed to confirm the planetary existence \citep{Wang2020, Zhou2023}. Such planets that do not host circumplanetary disks are normally expected to be only visible in total intensity \Itot\ images, but not in polarized \Qphi\ images. While circumplanetary disks might be potentially detectable in polarized light observations  \citep[e.g., simulations in][]{szulagyi21}, there are no confirmed detections with current instruments yet. A detection only in total intensity and not in polarized light would be a direct evidence of a planet that is embedded in a circumstellar disk \citep[e.g.,][]{Currie22a}. With the two observational modes in this work, we can investigate the existence of such planets: when a planet is embedded, the polarization fraction value of the region where it resides should be smaller. What is more, we can use the polarized observations to trace the leading and trailing spirals of forming planets \citep[e.g.,][]{Hammond2023}.

With the polarization fraction maps in Fig.~\ref{fig-polfrac}, however, we cannot directly recover any of the existing claims in the data here. This could be due to several factors: first, existing claims are reported in different wavelengths, making them not necessarily visible in $K_s$-band. Second, even if existing claimed planetary objects are visible in $K_s$-band, they can be fainter during our observational epochs due to variability \citep[e.g.,][]{Sutlieff2023}. Third, even if they are visible, they might have moved behind the coronagraph during our observation (e.g., AF~Lep~b detected in \citealp{Franson23, DeRosa23, Mesa23}, yet not in \citealp{Nielsen2019}). Last but not least, with longer wavelengths in $K_s$-band than in $J$-/$H$-band and thus worse spatial resolution (i.e., $1.22\lambda/D$, see Fig.~\ref{fig-app-transmission} for the IRDIS filters), planetary signal would be apparently more spread out and embedded onto disks due to broadening PSFs, even if the former does not physically colocate with the latter. This PSF-broadening effect further prevents a proper separation of planetary and disk signals, since it spreads planetary signals and make them appear less evident in polarization fraction maps. In principle, using only total intensity observations, here we could use high-pass filters to recover the disk-embedded planets. However, such an approach may require extensive tuning of the filtering parameters, and the highly structured disk morphology in this study further prevents a proper categorization between planetary and localized disk signals.

\section{Polarization fraction and color}\label{sec:pol-frac}
\begin{figure*}[bht!]
\centering
	\includegraphics[width=\textwidth]{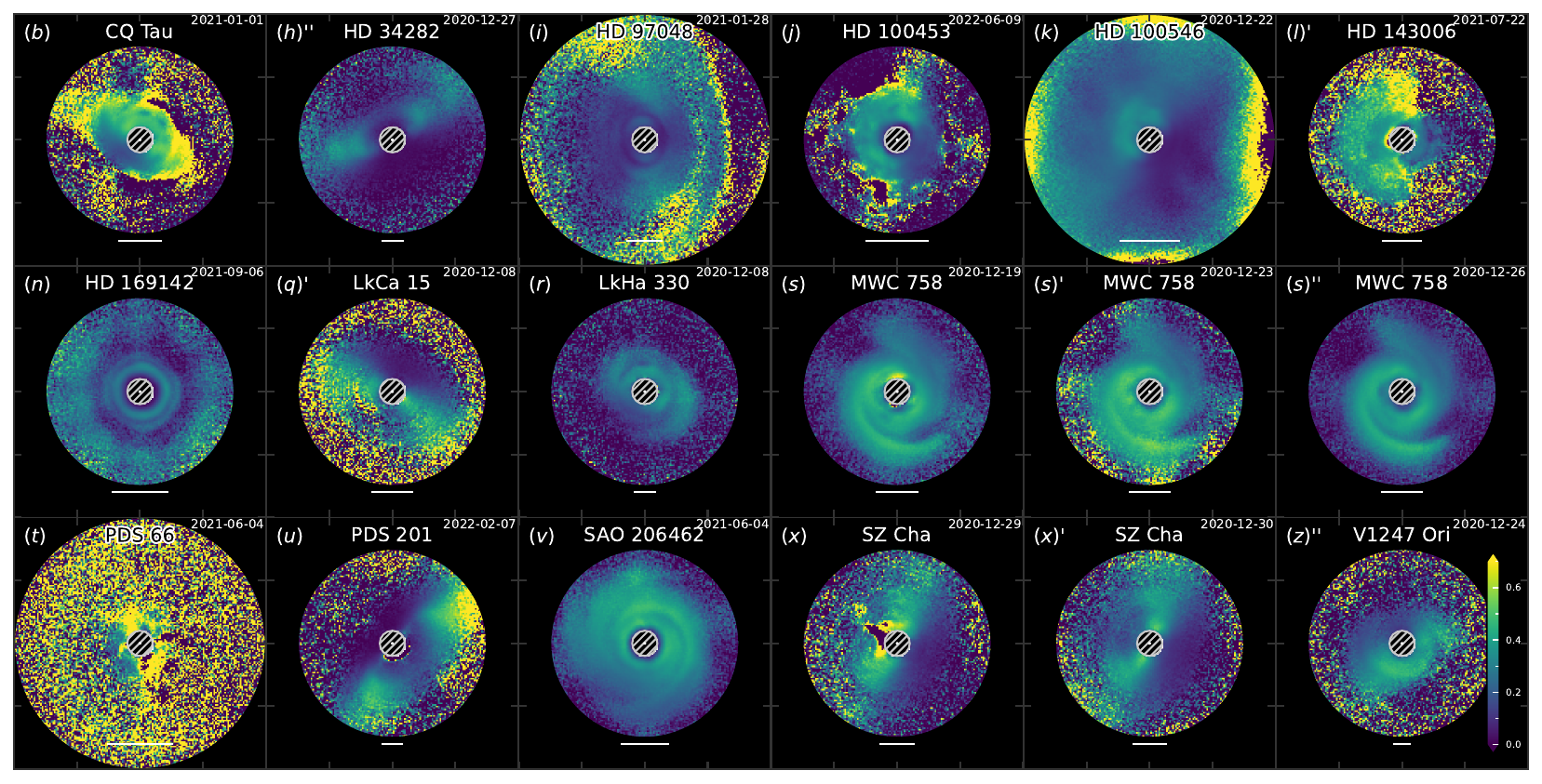}
    \caption{$K_s$-band polarization fraction maps with dimensions of $2\arcsec{\times}2\arcsec$ with identical color bars in linear scale, obtained from dividing the \Qphi\ maps in Fig.~\ref{fig-qphi} by the corresponding \Itot\ maps in Fig.~\ref{fig-tot}. \\ (The data used to create this figure are available.)}
    \label{fig-polfrac}
\end{figure*}

With the PDI \Qphi\ and RDI \Itot\ images, we calculated the polarization fraction maps by dividing them. The polarization fraction maps, also known as degree of linear polarization, are presented in Fig.~\ref{fig-polfrac}. To explore the ensemble properties of the scatterers, we study both their polarization fraction curves, which depict the polarization fraction dependence on the scattering angle, and their colors in polarized light.

\subsection{Parametric description}\label{sec:scaled-beta-1-component}

Polarization fraction curves can peak at different scattering angles and be asymmetric about the peak in both observations (e.g., \citealp{Munoz2021}, Fig.~6 therein; \citealp{Kiselev22}, Fig.~3 therein; \citealp{Frattin2022}) and theoretical studies (e.g., \citealp{Tazaki2019}, Figs.~2 and 4 therein; \citealp{chen20}, Fig.~17 therein). Motivated by the expectation that the polarization fraction is zero when the scattering angle is $0^\circ$ or $180^\circ$, we can describe a polarization fraction curve using a scaled beta distribution. For a scattering angle $\theta_{\rm scat} \in [0, \pi]$, which is the angle between the incident light vector and the scattered light vector, the polarization fraction follows

\begin{equation}\label{eq-pol-main-0}
f_{\rm pol}(\theta_{\rm scat}) \propto \theta_{\rm scat}^{\alpha-1}\left(\pi-\theta_{\rm scat}\right)^{\beta-1},
\end{equation}
where $\alpha > 1$ and $\beta > 1$. See Equation~\eqref{eq-pol-frac} in Appendix~\ref{sec-app-beta} for the full expression using scaled beta distribution in statistics, where we also used the peak polarization fraction parameter, $f_{\rm pol}^{\rm max}$, to control the maximum polarization fraction.

We implement the analytical polarization fraction curve for analysis. To depict the observed polarization fraction maps, we use the 3-dimensional symmetric and flared disk geometry in \texttt{diskmap}\footnote{\url{https://diskmap.readthedocs.io/en/latest/}} \citep{diskmap} by customizing its polarization fraction curve function. Specifically, \texttt{diskmap} can generate a 2-dimensional image from a 3-dimensional parameterized disk model. Even when some pixels have the same radial separation from the star, they can have different scattering phase angles in a flared disk. Using \texttt{diskmap}, we can convert between a total intensity image and a polarized image using a polarization fraction phase function. To implement this for our analysis, the vertical height of a disk in \texttt{diskmap} in cylindrical coordinates follows

\begin{equation}\label{eq-geometry-3d}
h(r) = h_0 \cdot \left(\frac{r}{1~{\rm au}}\right)^\gamma, 
\end{equation} 
where $h_0$ is the scale height and $\gamma$ is the flaring index, see Appendix~\ref{app-diskmap-geometry} for the details. For the 3-dimensional setup of the disk, we adopted the \citet{bohn22} results of the outer disks for both their inclination and position angle of the systems when available, and from \citet{Wagner2020} for PDS~201. In fact, we also fit the two angles independent of the values in previous publications, and did not obtain significant deviation for the position angle, neither did we have better constraints for the inclination of the systems. Thus, in order to reduce the computational cost with no loss of information, we adopted their published values.

To model the observed polarization fraction maps, we used the Savitzky--Golay filter in two-dimension\footnote{\url{https://github.com/espdev/sgolay2}\label{fn-sg}} to minimize the random noise in \Qphi\ data, see Appendix~\ref{app-sg-filter}. Specifically, based on the resolution in $K_s$-band, we used the Savitzky--Golay filter to remove the random noise for the observational data by fitting 5-degree polynomials with a 11-pixel window. In this way, we can fit smooth polarization fraction models to them without resolution degradation. Without the  Savitzky--Golay smoothing, the best-fit models are prone to shot noise, and do not produce consistent results across multiple observations even for a same system (e.g., SZ~Cha in Figs.~\ref{fig-app-pol-tot} and \ref{fig-app-pol-tot-frac}).

We explored the parametric dust scattering parameters in Equation~\eqref{eq-pol-main-0} and the peak polarization fraction $f_{\rm pol}^{\rm max}$, together with geometrical parameters (i.e., scale height $h_0$, flaring index $\gamma$) in Equation~\eqref{eq-geometry-3d}, using \texttt{emcee}\footnote{\url{https://emcee.readthedocs.io/en/stable/}} \citep{emcee} which performs Markov chain Monte Carlo (MCMC) exploration. We minimize the residuals by maximizing the following log-likelihood function, 

\begin{align*}\label{eq-loglike}
\ln\mathcal{L}\left(\bm{\Theta}\mid X_{\rm obs}\right) = &-\frac{1}{2}\sum_{i=1}^{N}\left(\frac{X_{{\rm obs}, i} - X_{{\rm model}, i}}{\sigma_{{\rm obs}, i}}\right)^2\\
	&- \sum_{i=1}^{N}\ln\sigma_{{\rm obs}, i} - \frac{N}{2} \ln(2\pi), \numberthis
\end{align*}
where $\bm{\Theta}$ denotes the set of parametric scattering and geometrical parameters in Equations~\eqref{eq-pol-main-0} and \eqref{eq-geometry-3d}, $\sigma_{\rm obs}$ is the uncertainty map for the polarization fraction map. In the above function, we also assume that the pixels $i$ follow independent normal distributions. The $X_{\rm obs}$ and $X_{\rm model}$ parameters denote the observation and model datasets for the polarization fraction maps, respectively.

To obtain the MCMC results, we limited the disk parameters $\bm{\Theta}$ in Equations~\eqref{eq-pol-main-0} and \eqref{eq-geometry-3d} using uniform priors, with $0 < h_0 < 0.2$, $0<\gamma<2$, $0\leq f_{\rm pol}^{\rm max}\leq 1$, $1< \alpha \leq5$ and $1< \beta \leq5$. Using the polarization fraction data from Fig.~\ref{fig-polfrac}, we present the best-fit polarization fraction curves in Fig.~\ref{fig-pol-frac} and the corresponding parameters to generate the curves in Table~\ref{tab:beta}. In Appendix~\ref{app-res}, we present the images for the corresponding models and residuals. For the $\alpha$ and $\beta$ parameters of the beta distribution, we limited the upper limit to $5$ since we do not observe significant curve change when we change the upper limit to larger values, since the internal data variation across different observations (see MWC~758 in Fig.~\ref{fig-pol-frac}) dominates the statistical variations. In addition, the \texttt{emcee} posteriors have extremely narrow ranges for the investigated parameters, and thus we do not present the corresponding ranges in Fig.~\ref{fig-pol-frac}, nor do we present the credible intervals in Table~\ref{tab:beta}.

\begin{figure}[h!] 	
\includegraphics[width=0.5\textwidth]{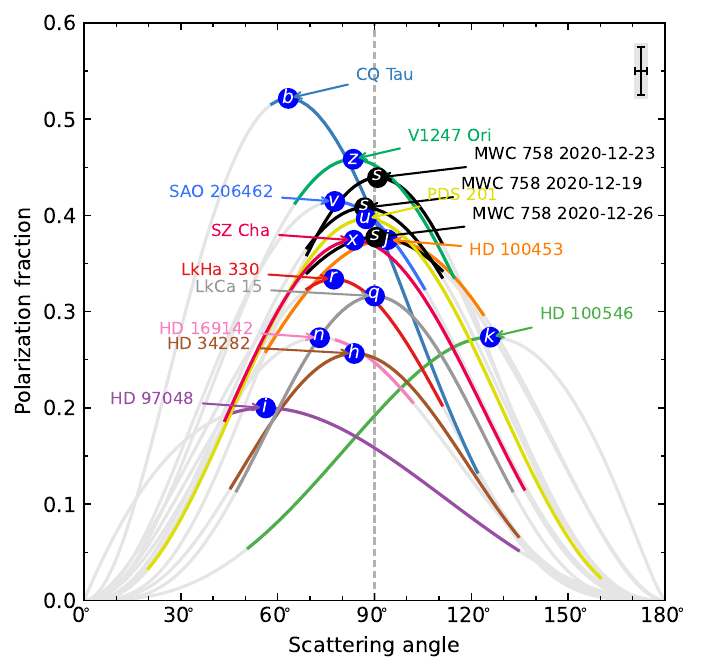}
    \caption{$K_s$-band polarization fraction curves, see Appendix~\ref{sec-app-beta} and Table~\ref{tab:beta} for the analytical expressions using scaled beta distribution. Best-fit curves with gray segments inaccessible from observation, based on system inclination assuming infinitely thin disks. The error bar on the top right, which could be used to infer the systematic uncertainty from the fitting method, is from the standard deviation of three sets of MWC~758 best-fit results.}
    \label{fig-pol-frac}
\end{figure}

\begin{figure*}[thb!] 	
\centering
\includegraphics[width=0.95\textwidth]{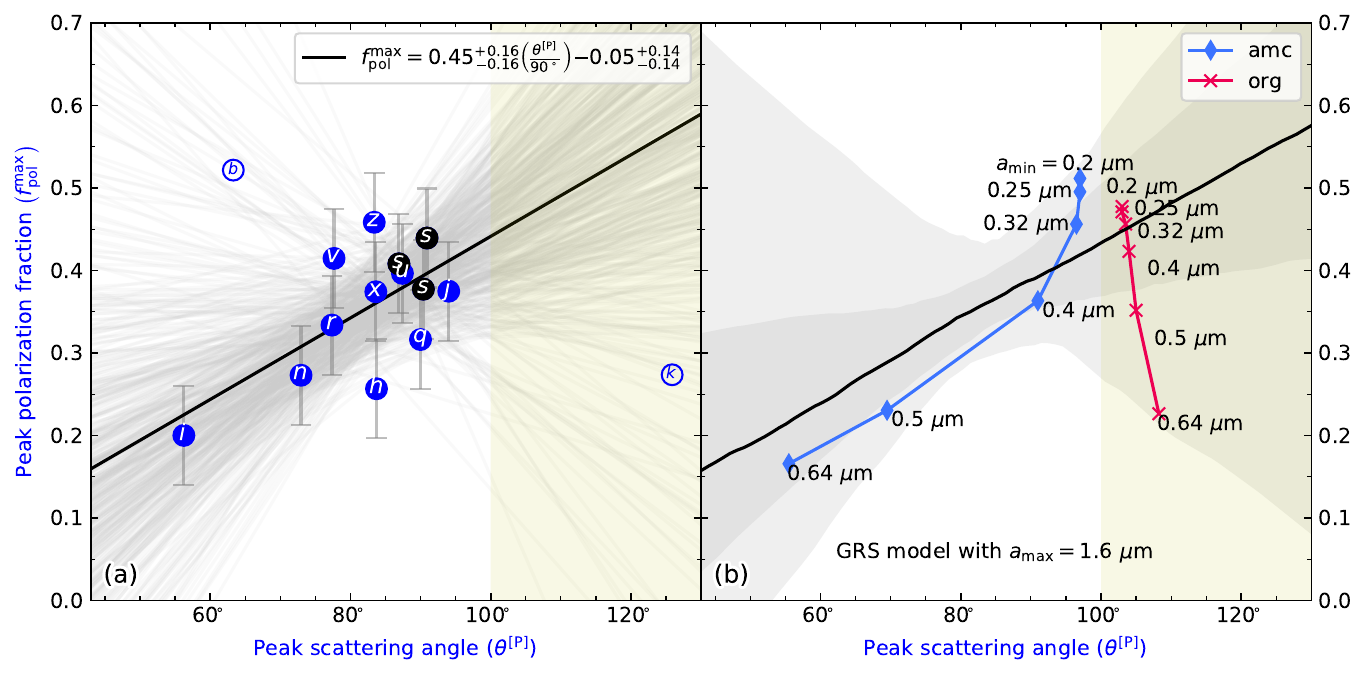}
    \caption{Scattering peak values in angle and fraction (SPAF) plot. (a) Tentative correlation between peak polarization fraction value and peak scattering angle, with the gray lines randomly fitting 6 to 12 systems for correlation exploration. The best-fit expression is performed on all data points here with a $0.06$ uncertainty for the peak polarization fraction, and certain systems (For $b$ or CQ~Tau and $k$ or HD~100546, marked with hollow symbols; for $l$ or HD~143006, a scaled beta distribution cannot describe its polarization fraction curve, potentially due to large scale shadowing in \citealp{Benisty2018}.) were excluded from Fig.~\ref{fig-pol-frac} to illustrate this relationship, see Sect.~\ref{sec-spaf-texts} for detailed discussion. (b) Gaussian random sphere (GRS) dust models from \citet{Tazaki22} and \citet{Tazaki23} by varying the minimum dust size for irregular compact grains, overlaid on the $1\sigma$ and $2\sigma$ ranges from the resampling results in (a). We observe a similar trend for absorptive material (``amc'') but not for less absorptive materials (``org''), with the latter neither providing peak scattering angles with the observations that are consistent beyond ${\sim}100^\circ$, see Sect.~\ref{sec-spaf-model-texts}. We note that the dust models do not necessarily reproduce the polarization fraction curves in Fig.~\ref{fig-pol-frac}.}
    \label{fig-pol-frac-modeling}
\end{figure*}

From an alternative approach, we also experimented forward modeling to obtain the polarization fraction curves. We first generated a total intensity disk model using \Qphi\ data using \texttt{diskmap}, then removed it from the original observations, and performed RDI reduction using KLIP \citep{soummer12} to minimize the residuals, see Appendix~\ref{sec:klipfm-q-tot-systems}. To reach high the computational efficiency and data quality, we do not adopt the KLIP forward modeling results, and instead present the results from direct polarization fraction modeling in Fig.~\ref{fig-pol-frac}.

With the best-fit 2-dimensional polarization fraction models, we notice that a one-component scaled beta distribution polarization fraction model in Equation~\eqref{eq-pol-main-0} cannot fully describe multi-component observations (especially, e.g., CQ~Tau, HD~100546, HD~143006, PDS~201), see Appendix~\ref{app-res}. In fact, one-component models are unable to capture local variations in smaller spatial scales (e.g., MWC~758, SAO~206462, V1247~Ori). Nevertheless, these models are still able to depict the large-scale variation even for spiral systems, especially in reproducing the regions with less polarization (i.e., north-west of MWC~758, south-east of SAO~206462) that might be otherwise categorized as other effects such as shadows.

Detailed inspection of the modeling residuals (e.g., Fig.~\ref{fig-polfrac-res}) suggests that the polarization fraction curves could vary within individual systems. For example, the multiple rings of SZ~Cha can have different polarization fraction curves. We focused only on the largest-scale structures for both ring and spiral systems here, to obtain a general understanding for the polarization fraction of protoplanetary disks. Future studies, including modeling the ring components separately for multi-ringed systems, and focused work on spiral systems, are necessary to quantify the difference in scattering properties (e.g., polarization fraction curve) within individual systems.

\subsection{Interpretation}

\subsubsection{Empirical trend}\label{sec-spaf-texts}
From the polarization fraction modeling results in Fig.~\ref{fig-pol-frac}, we observe that the peak polarization fraction is ${\lesssim}0.6$ for the systems in this study. In addition, the peak polarization fraction occurs at ${\lesssim}90^\circ$ scattering angle for nearly all systems. For certain systems, we could observe a tentative trend: the peak polarization fraction positively correlates with the peak scattering angle, see the scattering peak values in angle and fraction (SPAF) plot in Fig.~\ref{fig-pol-frac-modeling}(a), where we assigned an uncertainty of $0.06$ in peak polarization fraction for all samples (from a combination of a statistical uncertainty in the residual maps from modeling, with a systematic uncertainty from MWC~758 observations, see Fig.~\ref{fig-polfrac-res}). We did not observe a dependence of peak polarization fraction, or its corresponding scattering angle, on system inclination: Fig.~\ref{fig-pol-frac} is therefore not prone to modeling bias due to system inclination angles.

To generate the SPAF trend in Fig.~\ref{fig-pol-frac-modeling}(a), we did not include CQ~Tau, HD~100546, or HD~143006. The three excluded systems have more than two asymmetric disk components that are mutually superimposed in their polarization fraction maps, and the superimposition will lead to non-credible results if they are fitted using a one-component scaled beta distribution. Specifically, first, CQ~Tau is spatially more elongated along the north-east region than the south-west region, and thus a single scaled beta distribution might not be able to model such a polarization fraction map, yielding high polarization fraction at small scattering angles. Second, HD~100546 has a ``bright wedge'' in its south-west region \citep{Garufi2016}, and it is also observed in Fig.~\ref{fig-polfrac}, yielding high polarization fraction at large scattering angles. Third, HD~143006 has two ring components and large-scale self-shadowing effects \citep{Benisty2018}, and we observe that its polarization fraction along the north is significantly higher than that along the south, thus a one-component scaled beta distribution cannot describe its polarization fraction map. After all, if these three systems were included, the trend in Fig.~\ref{fig-pol-frac-modeling}(a) is less evident, and the inclusion of CQ~Tau would even make the trend a negative correlation.

With $13$ SPAF samples in Fig.~\ref{fig-pol-frac-modeling}(a), we notice a tentatively positive correlation between peak scattering angle and peak polarization fraction. To investigate the robustness of this SPAF correlation, we first randomly resampled 6 to 12 systems, then drew samples from normal distribution for the peak polarization fraction values, and performed linear fit. In this way, we can conservatively investigate the correlation for the entire population that are currently inaccessible. With dominating positive SPAF correlations from the resampling, we obtained negative correlation in ${\sim}10\%$ of this investigation. 

Given that ${\sim}10\%$ of the resampled trends have negative correlation, and that we have excluded systems for SPAF trend analysis (i.e., CQ~Tau, HD~100546, HD~143006), it is thus possible that positive SPAF trend in Fig.~\ref{fig-pol-frac-modeling}(a) could revert when more samples are available in the future. What is more, although the strong residuals in Fig.~\ref{fig-polfrac-res} (e.g., MWC~758, PDS~201) are included in the resampling exploration of Fig.~\ref{fig-pol-frac-modeling}(a), parameterizing the polarization fraction curves with scaled beta distributions could be limited. Moving forward, detailed dust model with different composition and geometry are needed to reproduce the samples in Fig.~\ref{fig-pol-frac-modeling}(a) and the observed polarization fraction curves in Fig.~\ref{fig-pol-frac}, especially when more observations are available, in the future.

\subsubsection{Dust model comparison}\label{sec-spaf-model-texts}

Polarization fraction could be indicative of dust properties, such as the porosity of aggregates and the size of their constituent grains \citep[i.e., monomers; ][]{Tazaki23}. With the extracted polarization fraction curves, on the one hand, the observed maximum polarization fraction of  ${\lesssim}0.6$ for the $K_s$-band in indicates that the monomers should be bigger than 100 nm (\citealp{Tazaki23}, Figure~6 therein). On the other hand, we can compare the observed polarization fractions with the predictions from radiative transfer numerical studies on different dust morphology.

To compare with numerical predictions, we first used the dust model results in $K_s$-band from the \texttt{AggScatVIR} database\footnote{\url{https://github.com/rtazaki1205/AggScatVIR}} \citep{Tazaki22, Tazaki23} and focused on irregular compact grains. We noticed that some of the predicted polarization fraction curves follow skewed bell-shaped profiles resembling the extracted curves in Fig.~\ref{fig-pol-frac}. Following the dust model categorization in \citet{Tazaki22}, we adopted the Gaussian random sphere (GRS) results therein, where there are two families of dust particles based on their composition. In each dust family, the constituting dust has a fixed mixture of four compositions (pyroxene silicate, water ice, carbonaceous material, and troilite) from \citet{Tazaki22}, see Table~1 therein for the optical constants. From the predictions, we extracted the maximum polarization fraction and its corresponding peak scattering angle for comparison with Fig.~\ref{fig-pol-frac-modeling}(a). 

The GRS model, namely the irregular compact grain, has a shape characterized by a power-law autocorrelation function with an index $\nu=-3.4$ and the relative standard deviation of the radius $\sigma=0.2$ \citep[see][for more detailed descriptions]{Nousiainen03}. Note that the GRS model is a single solid grain and, therefore, does not have a porosity or aggregate structure. The grains are assumed to obey a power-law size distribution with an index of $-3.5$ and the maximum grain radius of 1.6 $\mu$m. The minimum grain radius is a parameter of this study. For the two different compositions, on the one hand, in the Fig.~\ref{fig-pol-frac-modeling}(b) predictions, we observe that for absorptive materials (``amc''), when a maximum grain size is fixed to be $1.6~\mu$m, the peak polarization fraction decreases as the minimum dust size increases. This SPAF trend might be consistent with Fig.~\ref{fig-pol-frac-modeling}(a), with a caveat that the observed data are not representative of the SPAF population. On the other hand, however, such a SPAF trend cannot be reproduced using less absorptive materials (``org''). What is more, there could exist dust that are larger than $1.6~\mu$m that are accessible in $K_s$-band. After all, the GRS model might not be a representative description for dust in protoplanetary disks, and more careful modeling is needed to explain the observed polarization fraction for the systems in this study.

To explore beyond the GRS models, using the entire \texttt{AggScatVIR} database which also includes different dust properties (e.g., fractal/compact aggregates) at various porosity levels, we noticed that the peak value decreases with either increasing the dust radius or decreasing porosity. To explain the relatively low polarization fractions, the dust radius and porosity can be  degenerate in reproducing certain polarization fraction curves, indicating the potential diversity of scatterers in these systems.  However, we emphasize that these models roughly only reproduce the peak polarization fraction dependence on peak scattering angle, but do not match the extracted individual polarization fraction curves. Specifically, the predictions could have multiple local maxima in the polarization fraction curves, yet the beta distribution can only allow one: this mismatch is a limitation for our parametric description, yet it is hidden in the large uncertainties and more complicated parameterization is needed to describe the polarization fraction curves from the \texttt{AggScatVIR} database. To explain the observed polarization fraction maps, first, detailed dust model with different composition and geometry, as well as modeling multiple scattering effects, are needed. Second, observationally, separating the contributions between dust surface density and dust scattering properties could reduce the degeneracy. Third but not least, adopting parametric polarization fraction curves beyond beta distribution would allow multiple local maxima that are suggested in numerical models.

The dust polarization inferred from our observations may differ from those seen in the IM Lup disk surface. \citet{Tazaki23} found that fractal aggregates having a fractal dimension of 1.5 (i.e., dust mass $m\propto a_c^{1.5}$ with $a_c$ being the characteristic radius of an aggregate) with monomer size $a_\mathrm{mon}=0.2~\mu\mathrm{m}$ in the IM Lup disk surface when observed in $H$-band. In comparison, their best-fitting aggregate model  would suggest $f_\mathrm{pol}^\mathrm{max}=0.83$ and $\theta_\mathrm{max}=89^\circ$ at $K_\mathrm{s}$-band. However, none of our disk samples show such a high level of polarization fraction. Fractal aggregates are naturally formed through hit-and-stick coagulation, which is expected to occur during the early phases of dust coagulation. However, these fractal aggregates may not be long-lived as they quickly grow into larger aggregates and settle into the midplane \citep{Dullemond05, Tanaka05}. Eventually, the surface of the disk is dominated by particles replenished through collisional fragmentation of dust particles. What is more, the IM Lup disk is a young Class II disk with an estimated age of ${\sim}1.1$ Myr \citep{Avenhaus18}, whereas our samples used to derive the polarization fraction are generally older than IM Lup. The observed differences may reflect different stages of collisional dust evolution in the disks. In fact, using physical and chemical modeling, \citet{Cleeves2016} found that even mm-sized dust particles in IM Lup are lofted on to disk surface. Therefore, the distribution of dust particles on disk surface in IM Lup is likely not representative of those in our targets here.

Studying individual polarization fraction curves may provide dust information from an experimental approach. While scaled beta distributions here cannot describe the negative polarization fraction for large scattering angles (small phase angles: \citealp[e.g.][]{Munoz2021, Frattin2022}) in experimental studies, these angles are less accessible due to the inclination of the disks here in Fig.~\ref{fig-pol-frac}. With laboratory measurements showing diverse polarization fraction curves \citep[e.g.,][]{Munoz2021, Frattin2022}, they will offer high-quality phase curves for comparison with the observed ones.

\subsection{Disk color in polarized light}\label{sec-color}
\begin{figure}[tb!] 	
\centering
\includegraphics[width=0.5\textwidth]{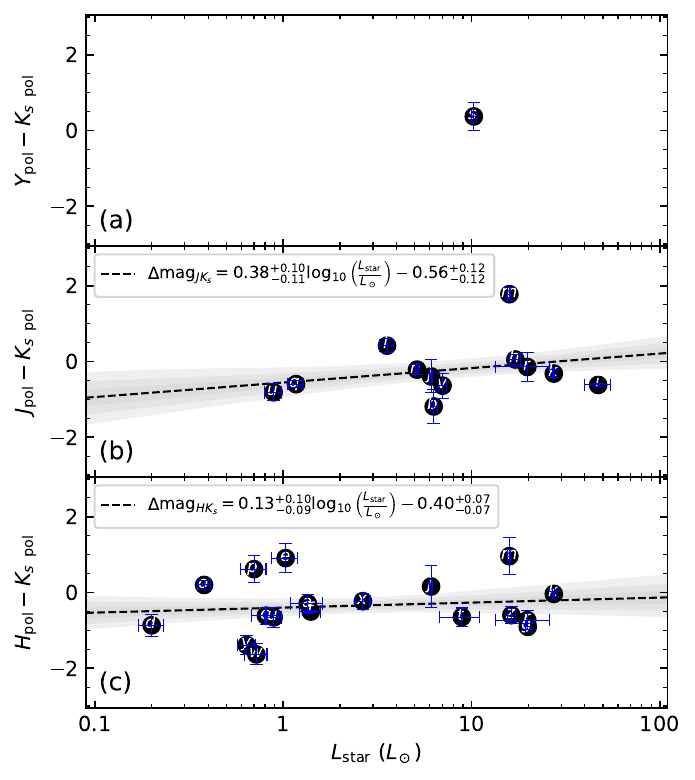}
     \caption{Polarized color at ${\approx}90^\circ$ scattering angle and stellar luminosity in Sect.~\ref{sec-color}. (a), (b), and (c) are the $Y$, $J$, and $H$-band data in polarized light in comparison with $K_s$-band data in polarized light, respectively. In $J-K_s$ comparison, increase in stellar luminosity leads to more neutral color; while the correlation is less evidence in $H-K_s$ comparison likely due to adjacent wavelengths (e.g., Fig.~\ref{fig-app-transmission}). Note: the bands are $1\sigma$, $2\sigma$ and $3\sigma$ confidence intervals from bootstrapping fit.\\ (The data used to create this figure are available.)}
    \label{fig-pol-color}
\end{figure}

From the color images, calculated by comparing $Y$-, $J$-, or $H$-band data with $K_s$ band data, we measured the colors at ${\sim}90^\circ$ scattering angle following \citet{Ren23}. We present in Fig.~\ref{fig-pol-color} the color dependence on stellar luminosity. Such a dependence can reflect the dust properties of the scatterers in these systems as well as limitations in the observations. 

We observe in Fig.~\ref{fig-pol-color}(b) that the observed protoplanetary disks are  blue in $J$ and $K_s$ bands for stars that are less luminous than ${\sim}10~L_\odot$. In $H$ and $K_s$ bands, the color of the disks can vary between blue and red. When stellar luminosity increases, we observe that both the $J-K_s$ and the $H-K_s$ color in polarized light are redder, and that the $H-K_s$ polarized color changes slower than that of $J-K_s$. The slow $H-K_s$ polarized color change is possibly caused by the adjacency of the two wavelengths in SPHERE/IRDIS, see Fig.~\ref{fig-app-transmission}. In comparison, \citet{crotts23} showed that debris disks in polarized light do also transition to redder color using the Gemini Planet Imager. The relatively redder color when stellar luminosity increases indicates that the scatterers are larger.

Measured at a ${\sim}90^\circ$ scattering angle, existing color studies in scattered light by \citet{Ren23} for debris disks showed a ubiquitously blue color in total intensity light by comparing between visible (${\sim}0.6~\mu$m) and the near-IR (${\sim}1.1~\mu$m and ${\sim}1.6\mu$m). With both studies having blue colors, we however note that the two studies are not comparable. The blue debris disks are between visible and near-infrared wavelengths (the latter is close to $J$/$H$ bands) in total intensity in \citet{Ren23}, while the protoplanetary disks here are between $J$- or $H$-band and $K_s$-band in polarized light.

\section{Summary}
\label{sec:summary}

We obtained $K_s$-band imaging of protoplanetary disks in scattered light using SPHERE/IRDIS on VLT for 29 systems in star-hopping mode. In the DPI setup of IRDIS imaging, we can obtain both polarized light observations and total intensity observations simultaneously.

By modeling the interior regions of the IRDIS $K_s$-band control ring using the information on the control ring with DI-sNMF, we have identified 15 systems in total intensity light with unprecedented data quality. For the RDI results from DI-sNMF, we calculated the companion detection limits for these observations with high-quality disk recovery: the existence of disks do raise the $K_s$-band detection limits in comparison to the exploration in $K_1$-/$K_2$-band in \citet{wahhaj21}. Nevertheless, an actual detection is a tradeoff between contrast and band-integrated companion luminosity, and thus narrower bands do not necessarily always provide better detections. Given that star-hopping observation has no dependence on sky rotation in the pupil-tracking mode, and that it can reach similar mass detection limits as ADI observations, it should be preferred to ADI observations in terms of observational schedulability.

Together with the IRDIS \Qphi\ data, we obtained the polarization fraction maps for these systems. With these polarization fraction maps, we can reduce the confusion by blob structures resembling planetary signals, since signals from giant protoplanets are not expected to be polarized. For the polarization fraction maps, we described the polarization fraction curves using analytical beta distributions. The polarization fractions peak between ${\sim}20\%$  and ${\sim}50\%$, yet they could be smaller than the actual values due to convolution effects from instrumentation. Assuming these polarization fraction curves are a credible representation of the actual polarization fractions, or if they undergo similar convolution effects, then we observe a tentative trend: the peak polarization fraction increases with the peak scattering angle. Using the \citet{Tazaki22} and \citet{Tazaki23} dust models from the \texttt{AggScatVIR} database, we could reproduce such a trend using absorptive materials for GRS dust; nevertheless, such models do not produce the individual polarization fraction curves. In addition, there can be alternative explanations with different dust parameters, and more future analysis and dust modeling are needed to interpret the observed polarization fraction curves. Moving forward, more comprehensive extraction of the polarization fraction curves -- including modeling the disk components separately -- can better help in comparing the scattering properties within each disk. In addition, lab measurements \citep[e.g.,][]{Munoz2021, Frattin2022} may provide important dust information for the observed polarization fraction curves.

For the 26 systems that have existing IRDIS observations in shorter wavelengths ($Y$-, $J$-, or $H$-band), we obtained the color of these systems at ${\sim}90^\circ$ scattering angle in polarized light. For $J_{\rm pol}-K_{s\ {\rm pol}}$ and $H_{\rm pol}-K_{s\ {\rm pol}}$ color in polarized light, we observe trends that the color is relatively redder when stellar luminosity increases. Such a trend indicates that the scatterers are larger for more luminous stars \citep[e.g.,][]{Ren23, crotts23}. In addition, while the polarized $H-K_s$ color here has a marginal trend of being relatively redder as stellar luminosity increases, the color ranges from red to blue for systems similar stellar luminosity, demonstrating the diversity of scatterers in different systems. In order to obtain the properties of the scatterers (e.g., mineralogy, morphology, porosity, size), detailed radiative transfer modeling efforts adopting realistic models \citep[e.g.,][]{Tazaki22, Tazaki23} are needed.

Using the SPHERE/IRDIS control ring for RDI data reduction with DI-sNMF, we cannot yet recover the disks in total intensity for systems with \textit{Gaia}~DR3 $R_p \gtrsim 11$ or 2MASS $K\gtrsim8$. For the sample with high selection bias here, our logistic regression results indicate that brighter hosts, redder references, and brighter references in observational wavelengths could aid in detecting disks. Given that there is no clear evidence that closer-in references can provide better RDI imagery for the hosts, star-hopping users can attribute a lower priority to on-sky proximity in reference selection. 

\begin{acknowledgements}
We thank the anonymous referee for their prompt and constructive comments. We thank Valentin Christiaens for comments on the manuscript. B.B.R.~thanks Yinzi Xin for discussions on wavefront sensing in high-contrast imaging, Jie Ma on convolution effects, and Laurent Pueyo for support. Based on observations collected at the European Organisation for Astronomical Research in the Southern Hemisphere under ESO programs \programESO{0103.C-0470}, \programESO{105.209E}, \programESO{105.20HV}, \programESO{105.20JB}, \programESO{106.21HJ}, and \programESO{108.22EE}. For the archival data in Sect.~\ref{sec-color}, based on observations collected at the European Organisation for Astronomical Research in the Southern Hemisphere under ESO programs \programESO{60.A-9389}, \programESO{60.A-9800}, \programESO{095.C-0273}, \programESO{096.C-0248}, \programESO{096.C-0523}, \programESO{097.C-0523}, \programESO{097.C-0702}, \programESO{097.C-0902}, \programESO{297.C-5023}, \programESO{198.C-0209}, \programESO{098.C-0486}, \programESO{098.C-0760}, \programESO{099.C-0147}, \programESO{0100.C-0452}, \programESO{0100.C-0647}, \programESO{0101.C-0464}, \programESO{0101.C-0867}, \programESO{0102.C-0162}, \programESO{0102.C-0453}, \programESO{0102.C-0778}, \programESO{1104.C-0415}, \programESO{0104.C-0472}, \programESO{0104.C-0850}, \programESO{109.23BC}, and \programESO{111.24GG}. This project has received funding from the European Research Council (ERC) under the European Union's Horizon 2020 research and innovation programme (PROTOPLANETS, grant agreement No.~101002188). This project has received funding from the European Union's Horizon Europe research and innovation programme under the Marie Sk\l odowska-Curie grant agreement No.~101103114. This work has made use of the High Contrast Data Centre, jointly operated by OSUG/IPAG (Grenoble), PYTHEAS/LAM/CeSAM (Marseille), OCA/Lagrange (Nice), Observatoire de Paris/LESIA (Paris), and Observatoire de Lyon/CRAL, and supported by a grant from Labex OSUG@2020 (Investissements d'avenir -- ANR10 LABX56). This research has made use of the SIMBAD database \citep{simbad}, operated at CDS, Strasbourg, France.  This research has made use of the VizieR catalogue access tool, CDS,  Strasbourg, France (DOI:  \href{https://doi.org/10.26093/cds/vizier}{10.26093/cds/vizier}). The original description of the VizieR service was published in A\&AS 143, 23 \citep{ochsenbein00}. The VizieR photometry tool is developed by Anne-Camille Simon and Thomas Boch. This research has made use of the Jean-Marie Mariotti Center \texttt{SearchCal} service\textsuperscript{\ref{fn-jmmc}} co-developed by LAGRANGE and IPAG.
\end{acknowledgements}

\bibliography{refs}

\appendix
\section{Auxiliary IRDIS imaging data}\label{app-pol-aux}

\begin{figure*}[htb!]
\centering
	\includegraphics[width=\textwidth]{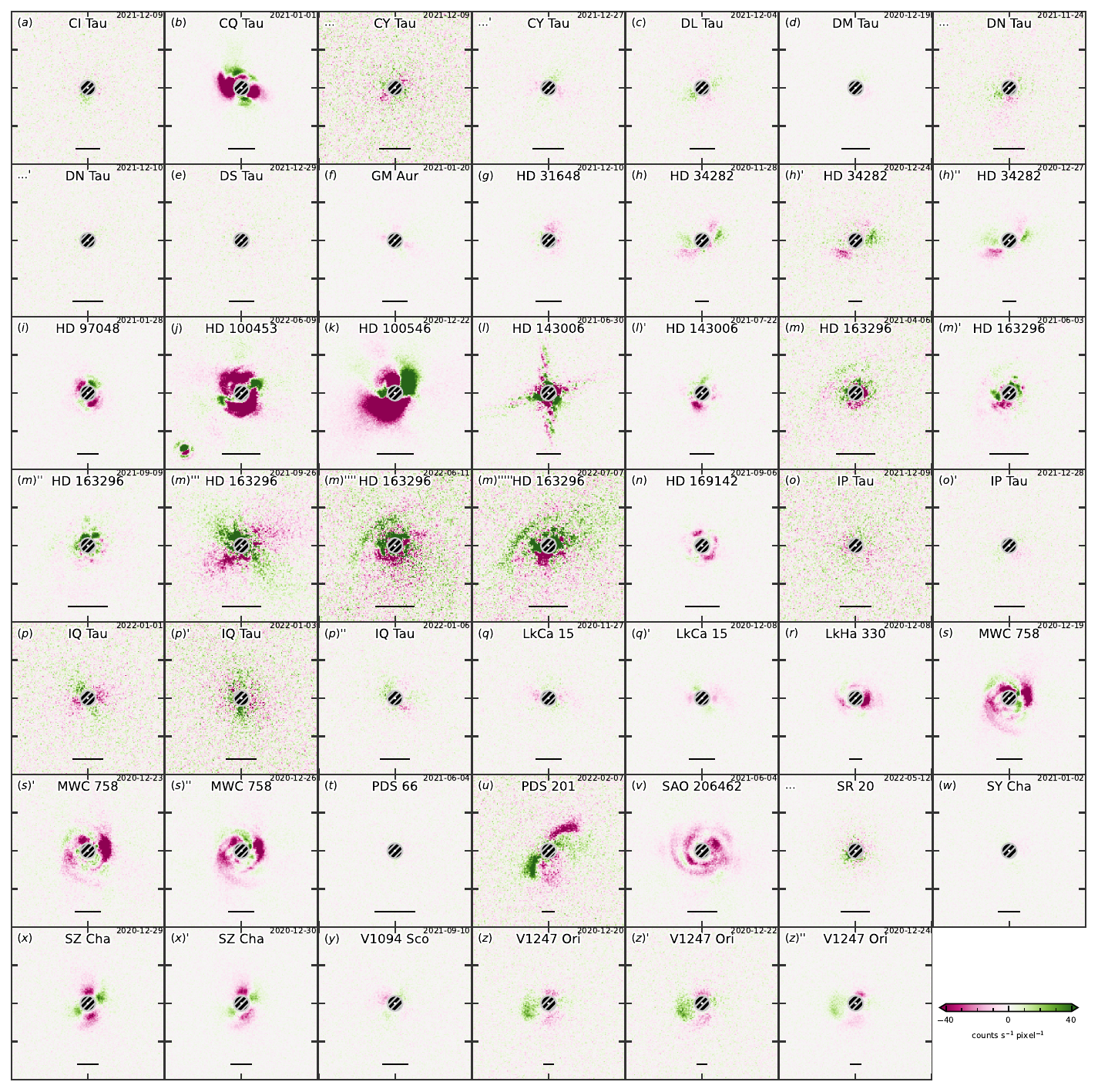}
    \caption{$K_s$-band \Uphi\ maps with dimensions of $2\arcsec{\times}2\arcsec$ with different color bars in linear scale. The field of view of each panel corresponds to those in Fig.~\ref{fig-qphi}. \\ (The data used to create this figure are available.)}
    \label{fig-uphi}
\end{figure*}

We present the stellar-signal-removed \Uphi\ images from \texttt{IRDAP} in $K_s$-band in Fig.~\ref{fig-uphi}. The absolute values of the \Uphi\ signals are ${\lesssim}5\%$ of the \Qphi\ signals, and thus the \Qphi\ images dominate the polarization signals for the protoplanetary disks in this study.

\begin{figure}[htb!]
\centering
	\includegraphics[width=0.45\textwidth]{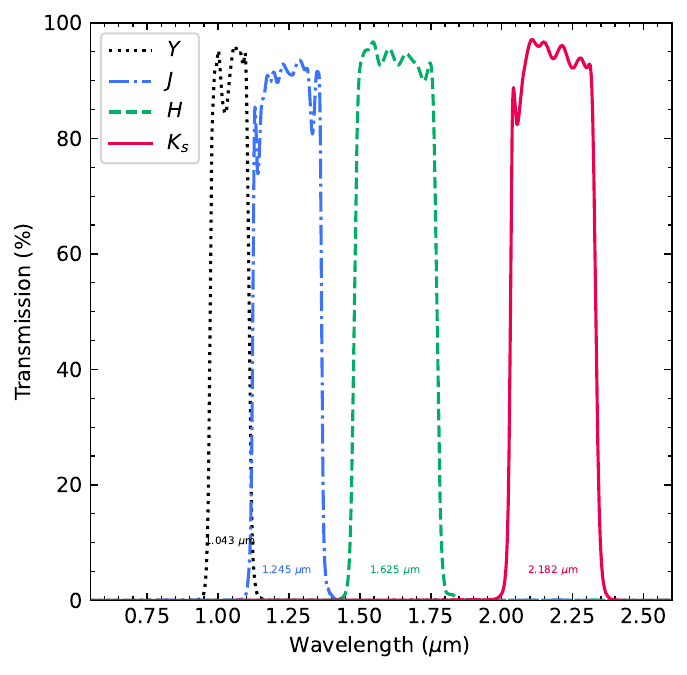}
    \caption{Transmission profiles, as well as the central wavelengths, for SPHERE/IRDIS in $Y$-, $J$-, $H$-, and $K_s$-band.}
    \label{fig-app-transmission}
\end{figure}

To study the \Qphi\ polarized color for the protoplanetary disks in this work, we summarize available SPHERE/IRDIS $Y$-, $J$-, and $H$-band observations in broadband polarized light in Table~\ref{tab:archive}, and compared them with the $K_s$-band data form this study. For HD~100546 and HD~163296 in $J$-band, we obtained the data from program 111.24GG and 109.23BC, respectively.  We show in Fig.~\ref{fig-app-transmission} the transmission profiles and central wavelengths\textsuperscript{\ref{fn-sphere-filters}} for the IRDIS filters with data analyzed in this study.

\begin{table*} 
\centering
\setlength{\tabcolsep}{5pt}
\caption{Archival SPHERE observations of $K_s$-band counterparts in $Y$, $J$, or $H$-band in polarized light \label{tab:archive}}
\begin{tabular}{cr | crr | crr | crr}    \hline\hline
id         & Target     & \multicolumn{3}{c|}{$Y$-pol} & \multicolumn{3}{c|}{$J$-pol} & \multicolumn{3}{c }{$H$-pol}\\
           &            & UTC        & $t^*_{\rm exp}$ & Program ID & UTC        & $t^*_{\rm exp}$ & Program ID & UTC        & $t^*_{\rm exp}$ & Program ID \\
           &            &            & (s)        &            & (s)        &            & (s)        \\
(1)        & (2)        & (3)        & (4)        & (5)        & (6)        & (7)        & (8)        & (9)        & (10)       & (11)       \\ \hline
\rowcolor{gray!10}$a$ & CI Tau     &            &            &            &            &            &            & 2017-12-07 & 576.0      & 0100.C-045 \\
\rowcolor{gray!10}  &            &            &            &            &            &            &            & 2018-01-02 & 2048.0     & 0100.C-045 \\
 $b$       & CQ Tau     &            &            &            & 2017-10-06 & 2000.0     & 098.C-0760 &            &            &            \\
           &            &            &            &            & 2018-02-18 & 1206.0     & 098.C-0760 &            &            &            \\
\rowcolor{gray!10}$\cdots$ & CY Tau     &             &            &            &            &            & & & &  \\ 
 $c$       & DL Tau     &            &            &            &            &            &            & 2018-11-15 & 480.0      & 0102.C-045 \\
           &            &            &            &            &            &            &            & 2018-12-29 & 301.5      & 0102.C-045 \\
           &            &            &            &            &            &            &            & 2018-12-29 & 640.0      & 0102.C-045 \\
\rowcolor{gray!10}$d$ & DM Tau     &            &            &            &            &            &            & 2018-10-02 & 384.0      & 0101.C-086 \\
\rowcolor{gray!10}  &            &            &            &            &            &            &            & 2018-10-02 & 3456.0     & 0101.C-086 \\
 $\cdots$  & DN Tau     &            &            &            &            &            &            & 2018-11-29 & 640.0      & 0102.C-045 \\
           &            &            &            &            &            &            &            & 2019-08-16 & 640.0      & 0102.C-045 \\
\rowcolor{gray!10}$e$ & DS Tau     &            &            &            &            &            &            & 2018-12-26 & 301.5      & 0102.C-045 \\
\rowcolor{gray!10}  &            &            &            &            &            &            &            & 2018-12-26 & 640.0      & 0102.C-045 \\
 $f$       & GM Aur     &            &            &            &            &            &            & 2018-01-02 & 2048.0     & 0100.C-045 \\
           &            &            &            &            &            &            &            & 2018-09-28 & 251.25     & 0101.C-086 \\
\rowcolor{gray!10}$g$ & HD 31648   &            &            &            &            &            &            & 2018-11-27 & 3328.0     & 0101.C-086 \\
 $h$       & HD 34282   &            &            &            & 2015-12-19 & 5632.0     & 096.C-0248 & 2017-12-08 & 64.0       & 60.A-9800  \\
           &            &            &            &            &            &            &            & 2017-12-08 & 64.0       & 60.A-9800  \\
           &            &            &            &            &            &            &            & 2017-12-08 & 804.0      & 60.A-9800  \\
           &            &            &            &            &            &            &            & 2017-12-08 & 670.0      & 60.A-9800  \\
           &            &            &            &            &            &            &            & 2017-12-08 & 2560.0     & 60.A-9800  \\
\rowcolor{gray!10}$i$ & HD 97048   &            &            &            & 2016-02-21 & 4096.0     & 096.C-0248 & 2017-05-14 & 256.0      & 099.C-0147 \\
 $j$       & HD 100453  &            &            &            & 2016-03-31 & 5376.0     & 096.C-0248 & 2018-06-05 & 2048.0     & 0101.C-046 \\
\rowcolor{gray!10}$k$ & HD 100546  &            &            &            & 2016-04-01 & 1280.0     & 096.C-0248 & 2019-02-18 & 1024.0     & 0102.C-016 \\
\rowcolor{gray!10}  &            &            &            &            & 2023-04-10 & 2304.0     & 111.24GG   & 2019-02-18 & 2688.0     & 0102.C-016 \\
 $l$       & HD 143006  &            &            &            & 2016-07-01 & 160.0      & 097.C-0902 &            &            &            \\
           &            &            &            &            & 2016-07-01 & 2560.0     & 097.C-0902 &            &            &            \\
\rowcolor{gray!10}$m$ & HD 163296  &            &            &            & 2016-05-26 & 2960.0     & 097.C-0523 & 2016-05-26 & 1024.0     & 097.C-0523 \\
\rowcolor{gray!10}  &            &            &            &            & 2023-06-09 & 2048.0     & 109.23BC   & 2016-05-26 & 512.0      & 097.C-0523 \\
 $n$       & HD 169142  &            &            &            & 2015-05-03 & 3200.0     & 095.C-0273 &            &            &            \\
           &            &            &            &            & 2015-07-03 & 16.0       & 60.A-9800  &            &            &            \\
\rowcolor{gray!10}$o$ & IP Tau     &            &            &            &            &            &            & 2019-11-24 & 3840.0     & 0104.C-085 \\
\rowcolor{gray!10}  &            &            &            &            &            &            &            & 2019-12-15 & 3584.0     & 1104.C-041 \\
 $p$       & IQ Tau     &            &            &            &            &            &            & 2018-12-19 & 301.5      & 0102.C-045 \\
           &            &            &            &            &            &            &            & 2018-12-19 & 640.0      & 0102.C-045 \\
\rowcolor{gray!10}$q$ & LkCa 15    &            &            &            & 2015-12-19 & 402.0      & 096.C-0248 &            &            &            \\
\rowcolor{gray!10}  &            &            &            &            & 2015-12-19 & 3840.0     & 096.C-0248 &            &            &            \\
 $r$       & LkHa 330   &            &            &            & 2017-10-06 & 2000.0     & 098.C-0760 & 2017-10-12 & 1920.0     & 0100.C-045 \\
           &            &            &            &            &            &            &            & 2017-12-06 & 1005.0     & 098.C-0760 \\
\rowcolor{gray!10}$s$ & MWC 758    & 2014-12-06 & 4096.0     & 60.A-9389  &            &            &            &            &            &            \\
\rowcolor{gray!10}  &            & 2015-03-04 & 3072.0     & 60.A-9389  &            &            &            &            &            &            \\
\rowcolor{gray!10}  &            & 2019-11-18 & 2560.0     & 0104.C-047 &            &            &            &            &            &            \\
 $t$       & PDS 66     &            &            &            & 2016-03-15 & 3072.0     & 096.C-0523 & 2016-03-16 & 3584.0     & 096.C-0523 \\
\rowcolor{gray!10}$u$ & PDS 201    &            &            &            &            &            &            & 2021-01-21 & 4672.0     & 1104.C-041 \\
 $v$       & SAO 206462 & 2015-03-31 & 728.625    & 095.C-0273 & 2015-05-03 & 4896.0     & 095.C-0273 &            &            &            \\
           &            &            &            &            & 2016-05-12 & 1024.0     & 097.C-0702 &            &            &            \\
           &            &            &            &            & 2016-06-22 & 2304.0     & 297.C-5023 &            &            &            \\
           &            &            &            &            & 2016-06-30 & 2304.0     & 297.C-5023 &            &            &            \\
           &            &            &            &            & 2022-03-30 & 2048.0     & 1104.C-041 &            &            &            \\
\rowcolor{gray!10}$\cdots$ & SR 20      &             &            &            &            &            & & & & \\ 
 $w$       & SY Cha     &            &            &            &            &            &            & 2017-05-16 & 2048.0     & 099.C-0147 \\
\rowcolor{gray!10}$x$ & SZ Cha     &            &            &            &            &            &            & 2017-03-21 & 2112.0     & 198.C-0209      \\
 $y$       & V1094 Sco  &            &            &            &            &            &            & 2017-03-12 & 2304.0     & 098.C-0486 \\
\rowcolor{gray!10}$z$ & V1247 Ori  & 2018-11-11 & 375.2      & 0100.C-064 &            &            &            & 2018-11-16 & 3456.0     & 0102.C-077 \\
\rowcolor{gray!10}  &            & 2018-11-11 & 750.4      & 0100.C-064 &            &            &            &            &            &            \\ \hline
\end{tabular}

\begin{flushleft}

{\tiny \textbf{Notes}: Column (1): Letter identifiers of the targets in this paper, the $\cdots$ symbols are used for systems with no existing polarized observations in other bands or without confident detection in $K_s$-band for polarized color extraction. Column (2): Target name. Columns (3), (6), and (9): UTC observation dates. Columns (4), (7), and (10): Total on-source exposure time for the target. Columns (5), (8), and (11): ESO Program ID. Different observation rows on the same observation night indicate different observation setups. For some observations, there were no PSF frames for relative flux measurement.
}

\end{flushleft}

\end{table*}

\section{Data deviation from NMF imputation}\label{app-di-ideal}
Missing data can impact the minimization of the cost function for matrix decomposition and dimensionality reduction methods. For NMF, \citet{ren20di} showed that the expected deviation due to missing data could follow a second-order form in their Equation~(33). Specifically, given a target image $T\in\mathbb{R}_{\geq0}^{1\times N_{\rm pix}}$ with $N_{\rm pix}\in\mathbb{N}$ pixels, with an NMF component basis vector $H_i\in\mathbb{R}_{\geq0}^{1\times N_{\rm pix}}$, we can denote the corresponding coefficient with $\omega_i\in\mathbb{R}_{\geq0}$. When a fraction of the data in $T$ is missing (or artificially ignored here), the corresponding coefficient is $\omega'_i\in\mathbb{R}_{\geq0}$. With these notations, Theorem~2 in \citet{ren20di} states that 

\begin{equation}
|\omega_i - \omega'_i| = o^2(\omega_i) = \omega_i \cdot o^2(1), \label{eq-omega-i-deviation-old}
\end{equation} where $o$ is the little-$o$ notation, meaning $|o(x)| \ll |x|$. Here we provide a derivation of the deviation under a more ideal assumption.\\

\noindent\textit{\textbf{Theorem~3} (Ideal Imputation). In the existence of missing data, if the cross-talk among NMF components is of the same order as the target modeling procedure, the influence of the missing data can reach a fourth-order deviation.}\\

\noindent\textit{\textbf{Proof.}} The second order deviation for Theorem 2 in \citet{ren20di} showed in their Equation~(27) that

 \begin{equation}
\omega_i = \frac{TH_i^T}{H_iH_i^T} \cdot \left(1+\sum_{j=1,j\neq i}^{n} \frac{\omega_j}{\omega_i} \cdot \frac{H_jH_i^T}{H_iH_i^T}\right)^{-1}, \label{eq-omega-i}
\end{equation}
has a second-order deviation for the multiplicand and the multiplier on the right-hand side. Therefore, the multiplication of two second-order terms would result into a second-order deviation. 

Focusing on the second term in the summand above, and following the same derivation process as Equation~(33) in \citet{ren20di}, we can obtain a second-order deviation for the cross-talk terms of $H$ in the second multiplier without loss of generality. Similar as Equation~\eqref{eq-omega-i-deviation-old}, we have

\begin{equation}\label{eq-cross-talk}
\left|\frac{H_jH_i^T}{H_iH_i^T}-\frac{(H_j\circ \mathbbm{1}_T)(H_i\circ \mathbbm{1}_T)}{(H_i\circ \mathbbm{1}_T)(H_i\circ \mathbbm{1}_T)^{T}}\right| = o^2\left(\frac{H_jH_i^T}{H_iH_i^T} \right),
\end{equation}
where $\mathbbm{1}_T\in\mathbb{B}^{1\times N_{\rm pix}}$ is an indicator matrix (in which $\mathbbm{1}_{Tj}=0$ when the corresponding element in $T_j$ is missing, and $\mathbbm{1}_{Tj}=1$ otherwise) which matches the dimension of $T$ and $H_i$.

Comparing the target modeling terms in Equation~(33) of \citet{ren20di}, i.e.,

\begin{equation}\label{eq-projection}
\left|\frac{TH_i^T}{H_iH_i^T}-\frac{(T\circ \mathbbm{1}_T)H_i^T}{(H_i\circ \mathbbm{1}_T)(H_i\circ \mathbbm{1}_T)^{T}}\right| = o^2\left(\frac{TH_i^T}{H_iH_i^T} \right),
\end{equation}
with the component cross-talk term in Equation~\eqref{eq-cross-talk}, if the target modeling term, $\frac{TH_i^T}{H_iH_i^T}$, and the cross-talk term among the components, $\frac{H_jH_i^T}{H_iH_i^T}$ $\forall j \neq i$, have identical orders of magnitude, we can rewrite Equation~\eqref{eq-omega-i} with missing data,

\begin{align*}
\omega'_i &= \frac{TH_i^T}{H_iH_i^T}\left[1-o^2(1) \right] \cdot  \left(1+\sum_{j=1,j\neq i}^{n} \frac{\omega_j}{\omega_i} \left\{\frac{H_jH_i^T}{H_iH_i^T} \left[1-o^2(1) \right] \right\} \right)^{-1} \numberthis \label{eq-omega-i-deviation-assumption-1} \\
		&= \frac{TH_i^T}{H_iH_i^T}\left[1-o^2(1) \right] \cdot \left(1+\sum_{j=1,j\neq i}^{n} \frac{\omega_j}{\omega_i} \frac{H_jH_i^T}{H_iH_i^T} \right)^{-1} \left[1+o^2(1) \right] \numberthis \label{eq-omega-i-deviation-assumption-2}\\ 
		&= \frac{TH_i^T}{H_iH_i^T} \cdot \left(1+\sum_{j=1,j\neq i}^{n} \frac{\omega_j}{\omega_i} \frac{H_jH_i^T}{H_iH_i^T}\right)^{-1} \cdot \left[1-o^4(1) \right] \\
		&= \omega_i \cdot \left[1-o^4(1) \right], \numberthis \label{eq-omega-i-deviation-new}
\end{align*}
which is a fourth-order deviation under small number approximation. To reach Equation~\eqref{eq-omega-i-deviation-new}, we applied the assumption of similar orders (i.e., ideal imputation requirement) between Equation~\eqref{eq-omega-i-deviation-assumption-1} and Equation~\eqref{eq-omega-i-deviation-assumption-2}.

In reality, the ideal imputation condition is not always guaranteed since the matrix elements are not equally contributing to the calculation. Therefore, following the same argument as Equation~(23) of \citet{ren20di}, the deviation introduced by missing data is between second and fourth order. \QED

The fourth-order deviation in Equation~\eqref{eq-omega-i-deviation-new} can be observationally approached in reality, as has been supported in this study in Fig.~\ref{fig-tot}. What is more, the recovery of the protoplanetary disks in total intensity from our study demonstrates that the control ring of SPHERE's adaptive optics system in an exposure is well-correlated with the interior PSF \citep[e.g.,][Fig.~2 therein]{Guyon21}. This demonstrates that we can use the control ring to infer the PSF interior to it, thus strategically avoiding the problem of overfitting that has been plaguing the post-processing of high-contrast imaging observations in total intensity. 

In this study, the usage of the SPHERE control ring has yielded beyond state-of-the-art results for the majority of the detected systems. Nevertheless, it is still limited by not only the existence of control rings, but also the objects of interest not superimposed on the control rings. We leave the handling of these limitations for future engineering \citep[e.g.,][]{Guyon21} and methodological studies for potential joint work.

\section{Parametric polarization fraction}\label{app-pol-total-klip}
To convert polarized light observations of disk-only signals to total intensity, existing studies adopted a bell-shaped polarization curve \citep[e.g.,][]{Engler2017, Olofsson18, lawson22}, with the curve physically motivated under the Rayleigh polarization regime. Specifically, by dividing the stellar-signal-removed local Stokes \Qphi~data from \texttt{IRDAP} \citep{irdap1, irdap2} by a polarization fraction map, one can convert polarized data to expected total intensity data. Combining this with a physically flared three-dimensional disk geometry (e.g., \texttt{diskmap}: \citealp{diskmap}), we should in principle obtain a well-described total intensity image from polarized light observations. 

To explore beyond the limitations from Rayleigh scattering \citep[e.g.,][]{ren21}, a scattering mechanism which is nevertheless valid only when dust particles are smaller than observation wavelength by more than one order of magnitude and the peak polarization is at $90^\circ$ scattering angle, we here adopt a parametric approach for extracting the best-fit polarization fraction curve. We obtain the polarization fraction by comparing \texttt{IRDAP} \Qphi\ data with total intensity data using \texttt{diskmap} while adopting an axisymmetric geometry for a flared disk. 

\subsection{Polarization fraction curve: scaled beta distribution}\label{sec-app-beta}
To address the fact that polarization fraction curves do not have to be symmetric around or peak at a scattering angle of $\frac{\pi}{2}$ \citep[e.g., Figure~17 of][]{chen20}, here we adopt  a parametric description of polarization fraction. For a scattering angle $\theta_{\rm scat} \in [0, \pi]$, the polarization fraction in Equation~\eqref{eq-pol-main-0} is

\begin{equation*}
f_{\rm pol}(\theta_{\rm scat}) \propto \theta_{\rm scat}^{\alpha-1}\left(\pi-\theta_{\rm scat}\right)^{\beta-1}.
\end{equation*}

In statistics, the probability density function (PDF) of a beta distribution follows,

\begin{equation}
B(x\mid \alpha, \beta) =  \frac{\Gamma(\alpha+\beta)}{\Gamma(\alpha)\Gamma(\beta)} x^{\alpha-1}(1-x)^{\beta-1},\label{eq-beta-og}
\end{equation}
for $x\in[0, 1]$, and $\Gamma(\cdot)$ is the gamma function with $\Gamma(x) = \int_{0}^{\infty}t^xe^{-t}~dt$ for $\forall x\in\mathbb{R}^+$.  We can normalize Equation~\eqref{eq-pol-main-0} to have $\frac{\theta_{\rm scat}}{\pi}$ follow a beta distribution form,

\begin{equation*}
f_{\rm pol}\left(\theta_{\rm scat} \mid \alpha, \beta\right) = \frac{1}{\pi^{\alpha+\beta-2}} \frac{\Gamma(\alpha+\beta)}{\Gamma(\alpha)\Gamma(\beta)} \theta_{\rm scat}^{\alpha-1}\left(\pi-\theta_{\rm scat}\right)^{\beta-1}. 
\end{equation*}

To enable observational description of polarization fraction, we can set the maximum polarization fraction to be $f_{\rm pol}^{\rm max}\in [0, 1]$. We now have a polarization fraction curve of

\begin{equation}\label{eq-pol-frac}
f_{\rm pol}\left(\theta_{\rm scat}  \mid \alpha, \beta,  f_{\rm pol}^{\rm max}\right)  = f_{\rm pol}^{\rm max} \cdot \frac{1}{B\left(\frac{\alpha-1}{\alpha+\beta-2} ~\middle\vert~ \alpha, \beta \right)} \cdot B \left(\frac{\theta_{\rm scat}}{\pi} ~\middle\vert~ \alpha, \beta \right),
\end{equation}
where $B(x \mid \alpha, \beta)$ is the original beta distribution PDF evaluated at $x$ using Equation~\eqref{eq-beta-og}. We use this parametric description of the polarzation fraction curves in this study.

In the polarization fraction curve in Equation~\eqref{eq-pol-frac}, its second multiplier is the inverse of the beta distribution PDF evaluated at its mode of $\frac{\alpha-1}{\alpha+\beta-2}$ (i.e., where the polarization fraction peaks), and thus it is used to regulate the maximum polarization fraction to be $f_{\rm pol}^{\rm max}$ together the first multiplier. The $\alpha$ and $\beta$ parameters also control the spread of the phase function, in the sense that the variance of a beta PDF is $\frac{\alpha\beta}{(\alpha+\beta)^2(\alpha+\beta+1)}$, or $(8\alpha+4)^{-1}$ when $\beta=\alpha$. In this study, we have $\alpha>1$ and $\beta>1$ to avoid mathematical divergence of the polarization fraction at $\theta_{\rm scat}\in\{0, \pi\}$.

\subsection{Implementation: 3-dimensional geometry}\label{app-diskmap-geometry}
To generate a polarization fraction map, \texttt{diskmap} needs the specification of the scale height, the maximum polarization fraction, the position angle and inclination angle of the disk. For the disk scale height,

\begin{equation}\label{eq-sh}
h(r) = h_0 \cdot \left(\frac{r}{1~{\rm au}}\right)^\gamma, 
\end{equation} 
where $r\in\mathbb{R}^+$ is the stellocentric distance in the midplane of the disk, $h_0\in\mathbb{R}^+$ is the disk scale height at $1$~au, and $\gamma\in\mathbb{R}^+$ describes the flaring of the disk. 

For the position angle and inclination angle values of the systems, we adopt the outer disk information from \citet{bohn22} when these information are available therein. For the polarization fraction function, we use the parametric description in Equation~\eqref{eq-pol-frac}. To extract the polarization fraction curves, we use \texttt{emcee} \citep{emcee} to explore the parameters in Equations~\eqref{eq-pol-frac} and \eqref{eq-sh},

\begin{equation}
\Theta=\{h_0, \gamma, \alpha, \beta, f_{\rm pol}^{\rm max}\},\label{eq-param-disk-diskmap}
\end{equation}
which are related with disk polarization to generate total intensity disk images using polarized images. 

We can obtain the polarization fraction curve in direct polarization fraction map comparison or forward modeling. On the one hand, from a direct measurement approach, we use the data imputation results and directly compare them with the polarization fraction map models. On the other hand, from a forward modeling approach, for a given set of parameters, we subtract the corresponding total intensity model from the preprocessed data, then perform Karhunen--Lo\`eve image projection (KLIP; \citealp{soummer12, amara12}) data reduction. For both approaches, we distribute the calculations using the \texttt{DebrisDiskFM} \citep{ren19} framework to reduce real-time cost of parameter exploration on a computer cluster. We minimize the residuals to obtain the best-fit parameters while assuming the pixels are independent from each other. We present the best-fit profiles in Fig.~\ref{fig-pol-frac} from the direct measurement approach, with the corresponding values in Table~\ref{tab:beta}.

\begin{table*} 
\centering
\setlength{\tabcolsep}{22pt}
\caption{Best-fit scaled beta distribution description for polarization fraction curve for Fig.~\ref{fig-polfrac} \label{tab:beta}}
\begin{tabular}{crccrcc}    \hline\hline
id    & Target     & Date         & $\theta^{[P]}$  & $f_{\rm pol}^{\rm max}$ & $\alpha$   & $\beta$   \\
(1)   & (2)        & (3)          & (4)             & (5)                  & (6)        & (7)       \\ \hline
$b$   & CQ Tau     & 2021-01-01   & $63\fdg3$       & 0.522                & 3.170      & 5.000     \\
$h$   & HD 34282   & 2020-12-27   & $83\fdg7$       & 0.257                & 4.478      & 5.000     \\
$i$   & HD 97048   & 2021-01-28   & $56\fdg3$       & 0.200                & 2.012      & 3.227     \\
$j$   & HD 100453  & 2022-06-09   & $94\fdg0$       & 0.375                & 3.069      & 2.892     \\
$k$   & HD 100546  & 2020-12-22   & $125\fdg9$      & 0.274                & 4.035      & 2.305     \\
$n$   & HD 169142  & 2021-09-06   & $73\fdg0$       & 0.273                & 3.209      & 4.241     \\
$q$   & LkCa 15    & 2020-12-08   & $90\fdg0$       & 0.316                & 5.000      & 5.000     \\
$r$   & LkHa 330   & 2020-12-08   & $77\fdg4$       & 0.334                & 4.016      & 5.000     \\
$s$   & MWC 758    & 2020-12-19   & $86\fdg9$       & 0.408                & 3.576      & 3.759     \\
$s'$  & MWC 758    & 2020-12-23   & $90\fdg9$       & 0.439                & 4.362      & 4.293     \\
$s''$ & MWC 758    & 2020-12-26   & $90\fdg4$       & 0.377                & 2.806      & 2.790     \\
$u$   & PDS 201    & 2022-02-07   & $87\fdg4$       & 0.396                & 3.768      & 3.932     \\
$v$   & SAO 206462 & 2021-06-04   & $77\fdg7$       & 0.415                & 3.210      & 3.912     \\
$x$   & SZ Cha     & 2020-12-30   & $83\fdg6$       & 0.374                & 3.769      & 4.190     \\
$z$   & V1247 Ori  & 2020-12-24   & $83\fdg4$       & 0.458                & 3.329      & 3.698     \\ \hline
\end{tabular}

\begin{flushleft}

{\tiny \textbf{Notes}: The modeling results are obtained directly from modeling Fig.~\ref{fig-polfrac} in Section~\ref{sec:pol-frac}, instead of performing negative injection for KLIP RDI in Appendix~\ref{sec:klipfm-q-tot-systems}. Column (1): Letter identifiers of the targets in this paper. Column (2): Target name. Column (3): UTC observation date. Columns (4): Scattering angle with peak polarization. Columns (5), (6), and (7): maximum polarization fraction, and parameters used to generate the polarization fraction curves in Fig.~\ref{fig-pol-frac}, see Equation~\eqref{eq-pol-frac} for the mathematical profile using scaled beta distribution. In addition, we did not report the uncertainties from \texttt{emcee} modeling due to them being extremely small (see \citealp{wolff17} for a way to obtain more realistic uncertainties). With the values from Columns (5), (6), and (7), to generate an array of polarization fraction in Fig.~\ref{fig-pol-frac}, readers can use the following pseudocode with \texttt{scipy} \citep{scipy}: \texttt{$f_{\rm pol}^{\rm max}$*scipy.stats.beta.pdf($\theta_{\rm scat}/\pi, \alpha, \beta$)/scipy.stats.beta.pdf($\frac{\alpha-1}{\alpha+\beta-2}, \alpha, \beta$)}, where $\theta_{\rm scat}$ is an array of scattering angles in units of radians which is divided by $\pi$ so that $0\leq\theta_{\rm scat}/\pi\leq1$.}

\end{flushleft}

\end{table*}

\subsection{Experiment: KLIP forward modeling}\label{sec:klipfm-q-tot-systems}
\begin{figure*}[htb!]
	\includegraphics[width=0.5\textwidth]{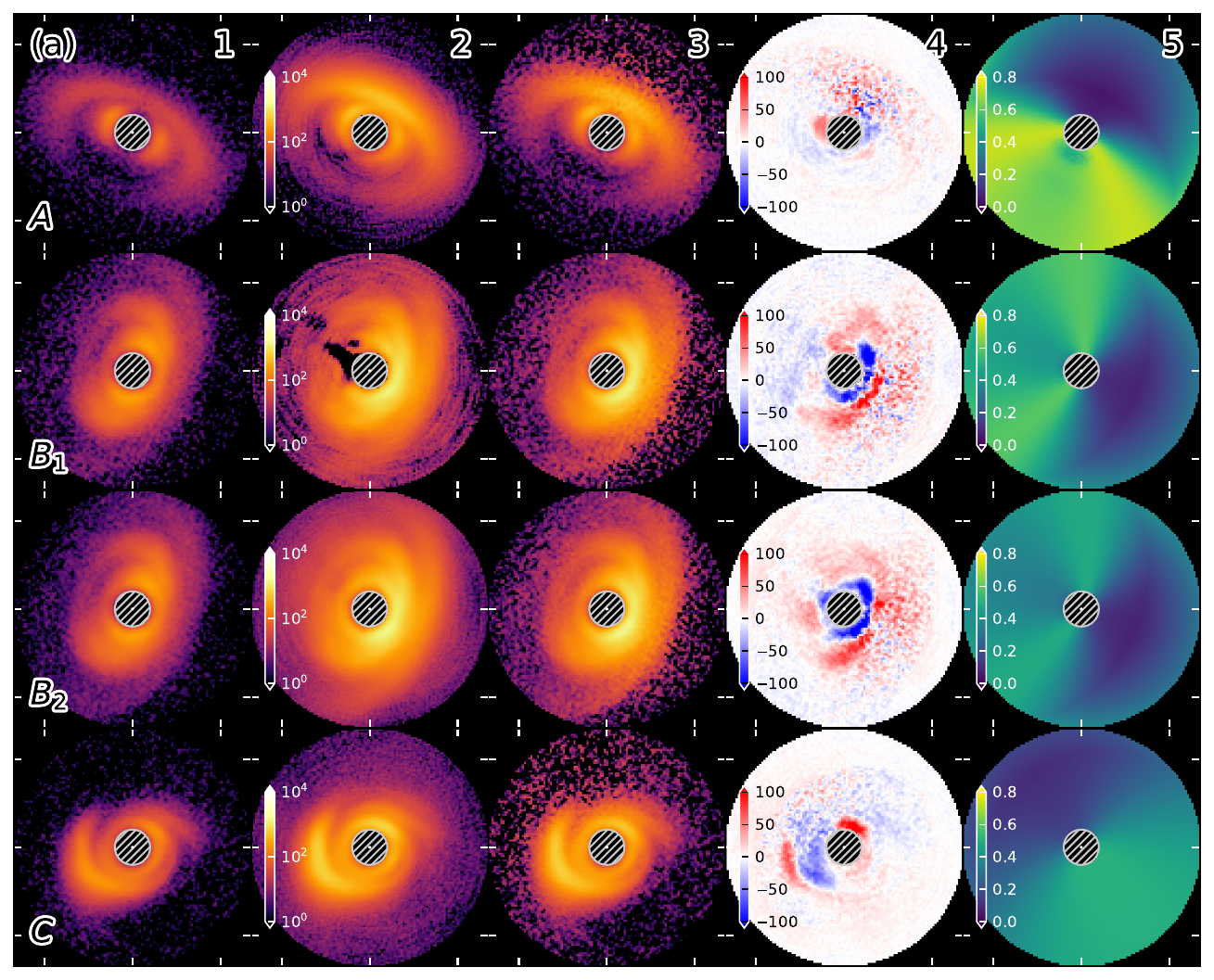}
	\includegraphics[width=0.5\textwidth]{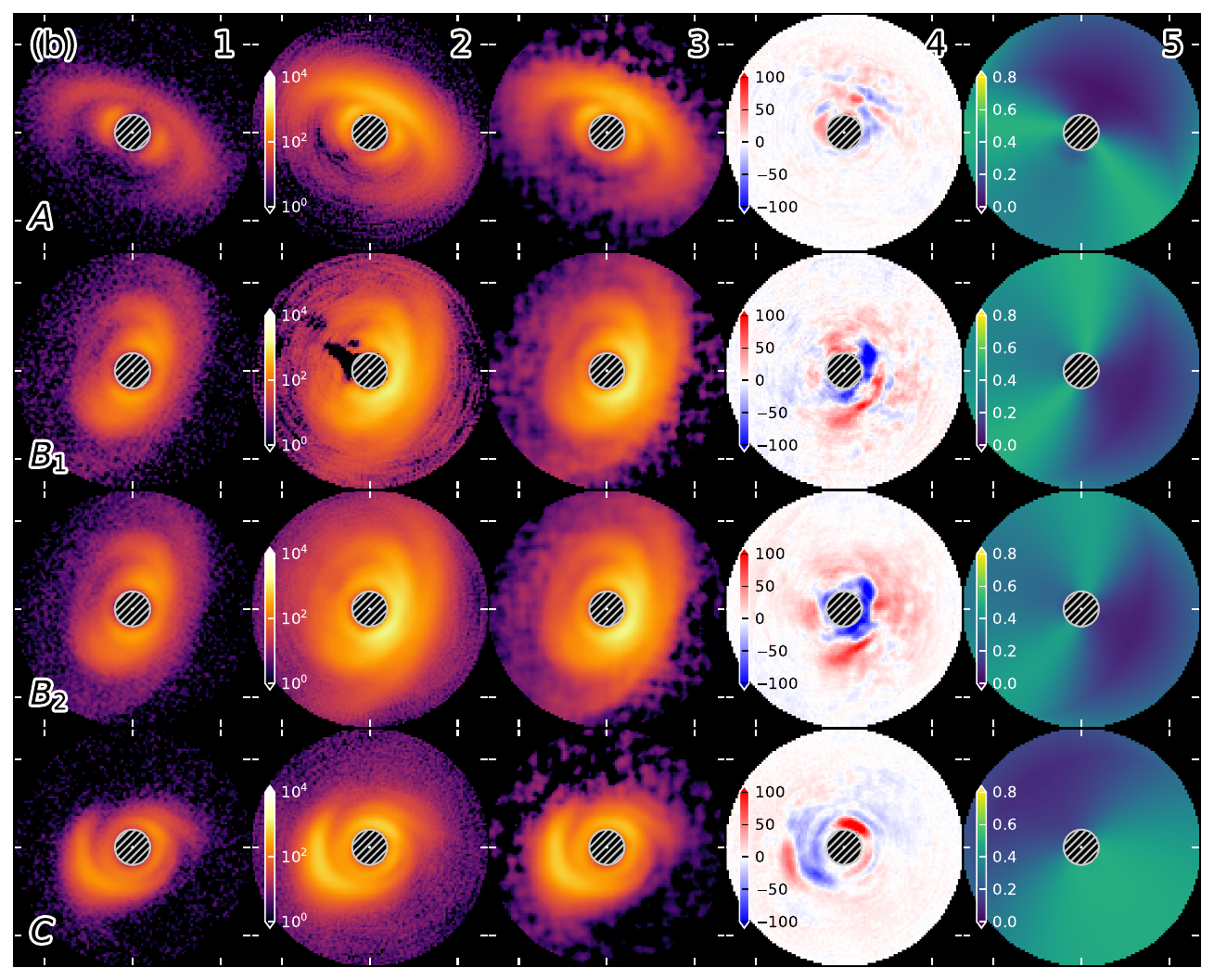}
    \caption{\Qphi\ to total intensity conversion using polarization fraction with scaled beta distribution, (a): direct polarization map division, (b): Savitzky--Golay filtered \Qphi\ data. From top to bottom, $A$: LkCa~15, on 2020 December 8; $B_1$: SZ~Cha, on 2020 December 29; $B_2$: SZ~Cha, on 2020 December 30; $C$: V1247~Ori, on 2020 December 24. From left to right,1: \Qphi\ data; 2: total intensity data from data imputation; 3: total intensity data converted from \Qphi\ data using polarization fraction maps; 4: RDI KLIP residuals using total intensity models; 5: polarization maps used to model total intensity data with RDI KLIP. A comparison between (a) and (b) shows an improvement of retrieving quality with the Savitzky--Golay filter. Nevertheless, the patterned residuals show that a single profile is limited in describing polarization fraction maps, especially when multiple disk components exist: modeling for separate disk components (e.g., rings in LkCa~15 and SZ~Cha) is needed for authentic description of polarization fraction curves.}
    \label{fig-app-pol-tot}
\end{figure*}

While we adopted the direct polarization fraction map modeling using the DI-sNMF results, KLIP has been the classical post-processing method in the high-contrast imaging of circumstellar structures. To study the application of scaled beta distribution polarization curve to KLIP, we also investigated the forward modeling approach to extract polarization fraction. Given that relatively simple geometry including ring structures can inform the three dimensional structures of protoplanetary disks in a straightforward way \citep[e.g.,][]{ginski16, deboer16}, we first applied the approach to ring systems. We then explored the applicability of the approach to spirals, see Figure~\ref{fig-app-pol-tot}(a) for the results for both morphologies and Figure~\ref{fig-app-pol-tot-frac}(a) for the best-fit polarization curves.

\subsubsection{Rings: LkCa~15 and SZ~Cha}
Using KLIP forward modeling, we present the results of LkCa~15 observed on UT 2020-12-08, SZ~Cha on UT 2020-12-29, and SZ~Cha on 2020-12-30. To produce Fig.~\ref{fig-app-pol-tot}, we removed a total intensity model from the observations, then performed KLIP data reduction to compare the results.

LkCa~15 hosts a two-ringed structure in scattered light \citep[e.g.,][]{Thalmann2016}. By directly dividing the \Qphi\ data by a polarization fraction map, the best-fit total intensity model resembles qualitatively the total intensity observation with data imputation. Nevertheless, the residual map shows pixelated islands, since the original \Qphi\ data has shot noise which do not fade away even after field rotation. In addition, although the outermost ring has a moderate level of residuals in Figure~\ref{fig-app-pol-tot}(a), strong residuals in the innermost ring suggests that the two rings have distinct polarization curves.

SZ~Cha hosts a three-ringed structure in scattered light. Similar as the two-ringed LkCa~15 disk, there exists pixelated residuals. With three rings in the system, the residuals are more evident, suggesting that the KLIP forward modeling approach is not able to depict the system with a simple flared disk assuming an identical polarization fraction curve. What is more, the recovered polarization fraction maps are at different brightness levels with different peak scattering angles in Figure~\ref{fig-app-pol-tot-frac}(a).

The pixelated residuals and excess disk residuals for the classical KLIP forward modeling approach suggests that the unmodified \Qphi\ data cannot be directly applied onto multi-ring systems with moderate inclinations. 

\subsubsection{Spirals: V1247~Ori}
For spirals we apply the KLIP forward modeling approach to the V1247~Ori data on 2020 December 24. Hosting at least a pair of spirals in scattered light \citep[e.g.,][]{Ohta2016}, the residual disk signals previous seen in ring systems are stronger for V1247~Ori. With the unideal experiments on ring systems, the unideal performance for the spiral system have been anticipated for the KLIP forward modeling approach.

\subsection{Limitations in polarization to total intensity conversion}
With the experiments above, in addition to existing evidences including the non-detection of polarized light in the most backward scattering regions for the HR~4796A system (e.g., polarized light: \citealp{perrin15}; total intensity: \citealp{milli17, ren20di}), we observe that the joint effect from scattering phase function and polarization fraction can result in the non-detection of polarized signals for certain regions that host total intensity signals. However, the non-detection of such polarized signals could not exclude their existence: there could exist such signals yet they are beyond the sensitivity limits of the existing instruments for given exposure times.

Given the facts that dust properties can vary as a function of stellocentric radius, that less efficient backward scattering could redistribute less light in observation, and that polarization fraction decrease can happen concurrently with backward scattering decrease, one should not expect to succeed in converting polarized images with no modifications to them to obtain perfect total intensity data for any disk system. 

As a potentially practical application, the method can be potentially applied to single-ring systems, and/or systems that have low inclinations. Even if this approach works, however, we note potential limitations including that the pixelated -- not smooth as the radiative transfer or simple geometric models generated in existing disk modeling work -- disk images in polarized light, when converted to total intensity light, can yield multi-modal distributions in the retrieved posteriors of the disk parameters in Equation~\eqref{eq-param-disk-diskmap}. While ignoring the existence of complex structures such as multiple rings or spirals, it is still necessary to remove the pixelated noise for the approach to work.

\subsection{Savitzky--Golay filter: advancing the conversion from polarization to total intensity}\label{app-sg-filter}
The pixelated residuals in Figure~\ref{fig-app-pol-tot}(a) should have originated from the rotation of the non-smooth pixelated data from \Qphi\ observations, and thus a smoothed version of the \Qphi\ data should be minimally adopted to minimize these residuals. In addition, a smoothing should not disperse the disk signals. Otherwise, smoothing would make it not directly comparable with the actual total intensity data, since the two observation modes are conducted at the same wavelengths on the same telescope instrument. 

The Savitzky--Golay filter was originally used to remove noise for one-dimensional data in \citet{SGfilter}. To smooth the \Qphi\ data for total intensity modeling, we use the Savitzky--Golay filter in two-dimension\textsuperscript{\ref{fn-sg}} to minimize the random noise in \Qphi\ data. In comparison with other classical convolution-based methods where signals are dispersed to remove noise, the Savitzky--Golay filter instead fit $p$-degree polynomials to data in moving windows. Motivated by the fact that the Rayleigh resolution for $K_s$-band with VLT/SPHRE is $69$~mas (i.e., $2.182~\mu$m for a $8.0$~m the telescope pupil seen by VLT/SPHERE), we perform $5$-degree polynomial for a window with 11-pixel (134.75~mas) width for the smoothing with Savitzky--Golay filter.

Using the Savitzky--Golay filtered \Qphi\ data, we reperform the study in Section~\ref{sec:klipfm-q-tot-systems} and present the images in Figure~\ref{fig-app-pol-tot}(b). In comparison with the results with the original \Qphi\ data in Figure~\ref{fig-app-pol-tot}(a), the Savitzky--Golay filtered data show smoother residuals, with the total intensity modeling being qualitatively more compatible with the total intensity observation using data imputation. Although there are still strong residuals for multi-ring or spiral systems, the Savitzky--Golay filter has removed the pixelated residuals in the original \Qphi\ data.

\begin{figure}[htb!]
	\includegraphics[width=0.5\textwidth]{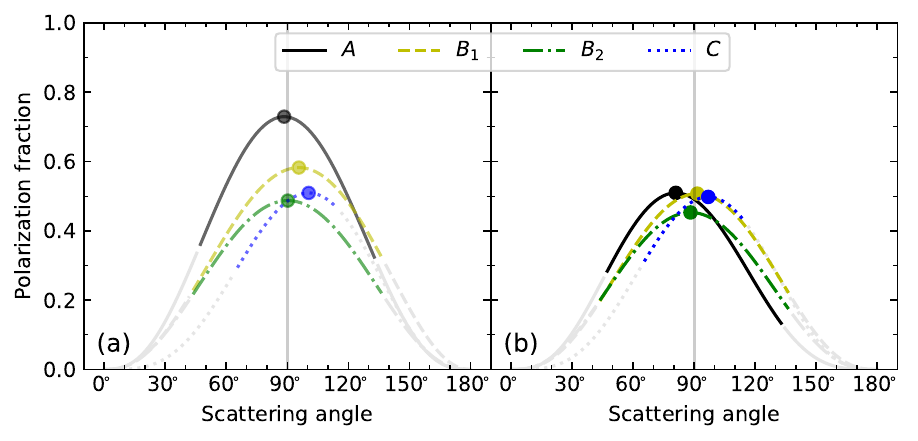}
    \caption{Best-fit polarization curves assuming beta distribution from RDI KLIP forward modeling in Figure~\ref{fig-app-pol-tot}, the peak polarization locations are marked with circles. The light gray curves denote scattering angle ranges that are not accessible for inclined thin disks with no flaring. (a) Direct conversion: the extracted profiles are different even for the same system (i.e., $B_1$ and $B_2$). (b) Conversion with Savitzky--Golay filtered \Qphi\ data: the extracted curves are less distinct, since such reductions are less prone to pixel-wise discrete noise.}
    \label{fig-app-pol-tot-frac}
\end{figure}

We present the extracted best-fit polarization fraction curves with KLIP forward modeling in Figure~\ref{fig-app-pol-tot-frac}. Without the Savitzky--Golay filter, the best-fit polarization fraction curves in Figure~\ref{fig-app-pol-tot-frac}(a) showed different profiles, yet such profiles are not valid since the extracted profiles are even distinct for the SZ~Cha data observed at different nights. The fact that the best-fit polarization fraction curves in Figure~\ref{fig-app-pol-tot-frac}(a) are different for SZ~Cha suggests that the original approach cannot be used to interpret or compare the profiles. With the Savitzky--Golay filter, the polarization fraction curves in Figure~\ref{fig-app-pol-tot-frac}(b) are less distinct from each other, and the similar profiles for SZ~Cha suggest that its distinct profiles in Figure~\ref{fig-app-pol-tot-frac}(a) are algorithmic effects.

For the similar profiles in different systems in Figure~\ref{fig-app-pol-tot-frac}(b), however, it is possible that the scattering angle and intensity corresponding to peak polarization fraction can still vary across different systems. Due to the assumption of independent pixels in Eq.~\eqref{eq-loglike}, the retrieved uncertainties are extremely small in Table~\ref{tab:beta} and thus were not presented. To further quantify the similarity or difference for the extracted profiles, proper uncertainty estimates should be performed \citep[e.g.,][]{wolff17}, and such estimates are beyond the scope of this study. 

For the polarization fraction modeling of high-contrast imaging observations, our experiment here shows that the usage of smoothed data is necessary to reduce bias for further comparison of different systems. To fully extract the profiles for comparison, more careful treatment of the correlated uncertainties are needed. In the main text of this study in Sect.~\ref{sec:analysis}, we obtain the polarization fraction profiles by directly modeling the polarization fraction maps generated from PDI \Qphi\ and DI-sNMF \Itot\ data. In addition, we used the results from three high-quality observations of MWC~758 in different nights, for a empirical estimation of the potential uncertainty of the polarization fraction curve modeling.

\section{Polarization fraction models}\label{app-res}
\begin{figure*}[bht!]
\centering
	\includegraphics[width=\textwidth]{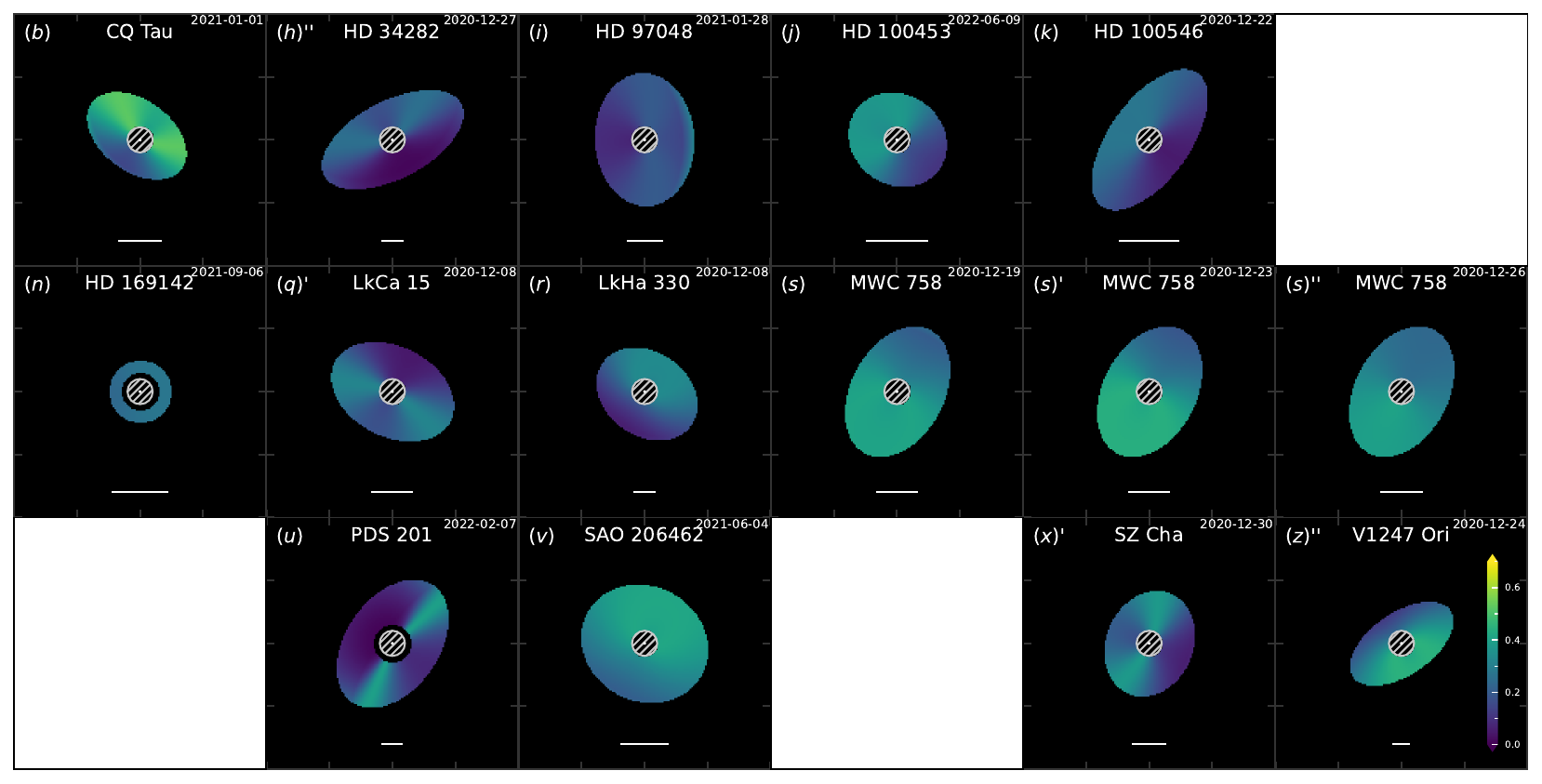}
    \caption{Models of $K_s$-band polarization fraction with dimensions of $2\arcsec{\times}2\arcsec$ with identical color bars in linear scale, see Fig.~\ref{fig-polfrac} for the observation. The non-masked areas are regions used for polarization fraction modeling with scaled beta distribution in Sect.~\ref{sec:pol-frac}. \\ (The data used to create this figure are available.)}
    \label{fig-polfrac-model}
\end{figure*}

\begin{figure*}[bht!]
\centering
	\includegraphics[width=\textwidth]{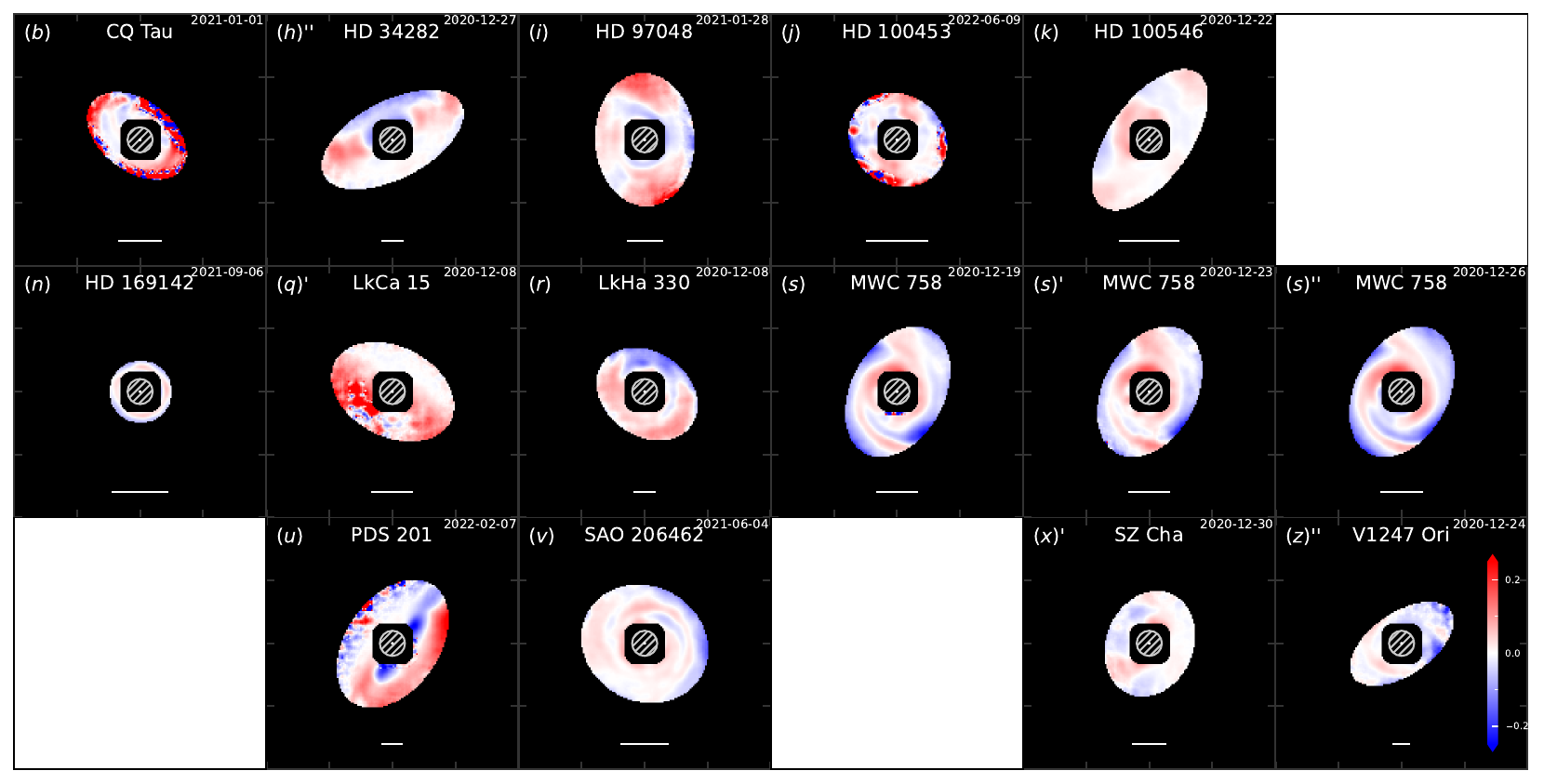}
    \caption{Residuals of $K_s$-band polarization fraction maps in Fig.~\ref{fig-polfrac} (smoothed by Savitzky--Golay filter in Sect.~\ref{app-sg-filter}) subtracted by the models in Fig.~\ref{fig-polfrac-model}, with dimensions of $2\arcsec{\times}2\arcsec$ with identical color bars in linear scale. Combining the standard deviation of each residual map of ${\approx}0.05$ statistically, with an uncertainty of ${\approx}0.03$ systematically from the three MWC~758 observations in Fig.~\ref{fig-pol-frac}, we assign a total uncertainty of $0.06$ for the maximum polarization fraction values.}
    \label{fig-polfrac-res}
\end{figure*}

We present the models and residuals for modeling the polarization fraction maps in Sect.~\ref{sec:pol-frac} here. Fig.~\ref{fig-polfrac-model} shows the regions and best-fit models using scaled beta distribution, and Fig.~\ref{fig-polfrac-res} contains the residuals by subtracting the models from Savitzky--Golay-smoothed observation in Fig.~\ref{sec:pol-frac}.

In the residual maps, we witness an ${\approx}0.1$ residual in polarization fraction, and certain levels of patterned residuals. The patterned residuals are due to the limitations in describing the polarization fraction curves using a single scaled beta distribution. In fact, multi-ringed systems, as well as non-ring systems such as spirals, likely have different scattering properties at different locations. This effect can be also seen in the residual maps in Fig.~\ref{fig-app-pol-tot}.

\section{Contrast curves from ADI}\label{app-adi-contrast}
\begin{figure*}[t!] 	
\includegraphics[width=0.5\textwidth]{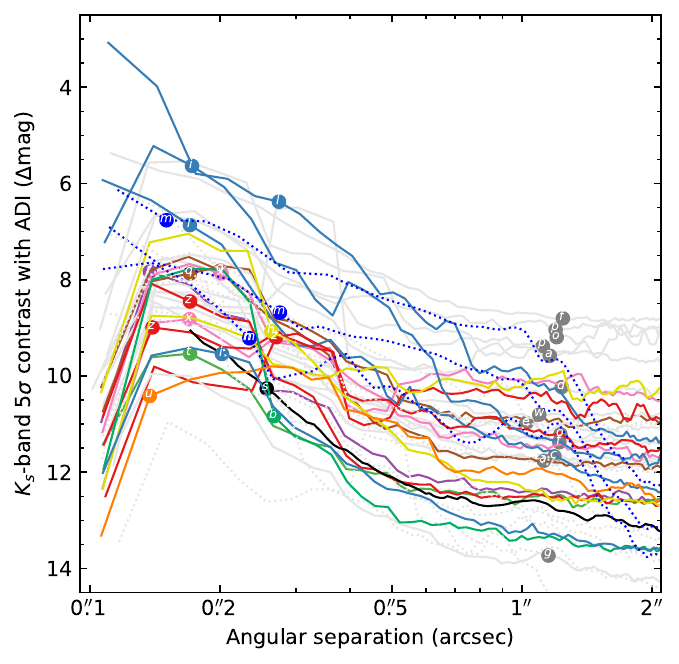} 
\includegraphics[width=0.515\textwidth]{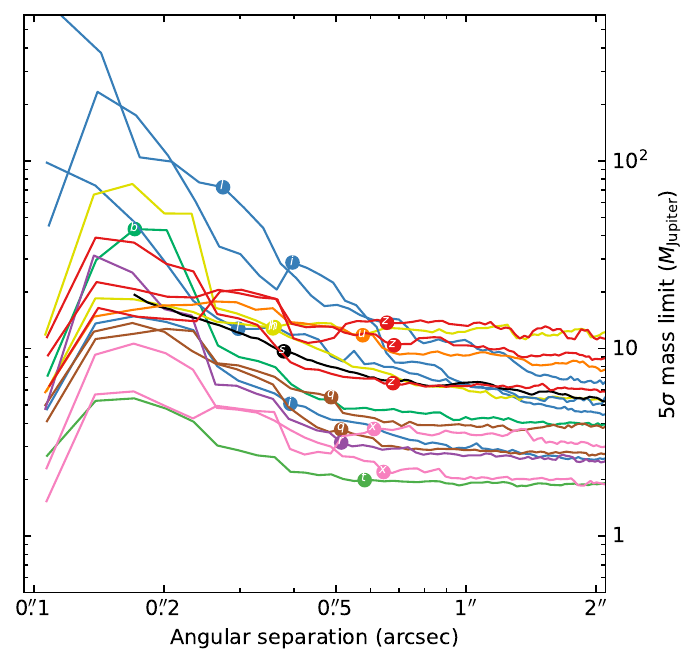} 
    \caption{ADI contrast curves obtained from High Contrast Data Centre (left) and AMES-cond mass limits (right) for systems in this study, the line colors are consistent with Fig.~\ref{fig-contrast}. Disk hosts with only PDI detections (or marginal PDI detections) are displayed with light gray color with solid lines, and annotated with gray symbol; reference stars are in dotted light gray lines. For HD~163296, the ADI contrast curves are displayed and annotated in blue. \\ (The data used to create this figure are available.)}
    \label{fig-contrast-adi}
\end{figure*}

We compared the RDI contrast curves from DI-sNMF with the RDI contrast curves from the High Contrast Data Centre and obtained consistent results. Using the High Contrast Data Centre products, we present the ADI contrast curves for the datasets presented in this work here. Specifically, we present the ADI contrast curves for both the target stars and their reference stars here; when the field rotation is not sufficient for ADI reduction, we present the non-ADI contrast curves. We compared the ANDROMEDA  \citep{andromeda1, andromeda2}, TLOCI \citep{tloci}, and KLIP \citep{soummer12, amara12} contrast curves, and present the deepest contrasts in Fig.~\ref{fig-contrast-adi}.

\end{CJK*}
\end{document}